# Biophysically detailed mathematical models of multiscale cardiac active mechanics

F. Regazzoni[*,1], L. Dedè[1], A. Quarteroni[1,2]

[1]*MOX - Dipartimento di Matematica, Politecnico di Milano,
P.zza Leonardo da Vinci 32, 20133 Milano, Italy*
[2]*Mathematics Institute, École Polytechnique Fédérale de Lausanne,
Av. Piccard, CH-1015 Lausanne, Switzerland (Professor Emeritus)*

**Abstract**

We propose four novel mathematical models, describing the microscopic mechanisms of force generation in the cardiac muscle tissue, which are suitable for multiscale numerical simulations of cardiac electromechanics. Such models are based on a biophysically accurate representation of the regulatory and contractile proteins in the sarcomeres. Our models, unlike most of the sarcomere dynamics models that are available in the literature and that feature a comparable richness of detail, do not require the time-consuming Monte Carlo method for their numerical approximation. Conversely, the models that we propose only require the solution of a system of PDEs and/or ODEs (the most reduced of the four only involving 20 ODEs), thus entailing a significant computational efficiency. By focusing on the two models that feature the best trade-off between detail of description and identifiability of parameters, we propose a pipeline to calibrate such parameters starting from experimental measurements available in literature. Thanks to this pipeline, we calibrate these models for room-temperature rat and for body-temperature human cells. We show, by means of numerical simulations, that the proposed models correctly predict the main features of force generation, including the steady-state force-calcium and force-length relationships, the length-dependent prolongation of twitches and increase of peak force, the force-velocity relationship. Moreover, they correctly reproduce the Frank-Starling effect, when employed in multiscale 3D numerical simulation of cardiac electromechanics.

# 1 Introduction

Cardiovascular mathematical and numerical models are increasingly used, with a twofold role [16, 25, 31, 32, 77, 96]. On the one hand, realistic and detailed in silico models of the heart can deepen the understanding of its function, help the interpretation of experimental observations and explain the delicate links between the organ-level emergent phenomena and the underlying biophysical mechanisms. On the other hand, patient-specific numerical simulations, which are increasingly becoming available, can provide clinicians with valuable quantitative information to improve patient care and to support decision-making procedures.

---

[*]Corresponding author. Email address: `francesco.regazzoni@polimi.it`

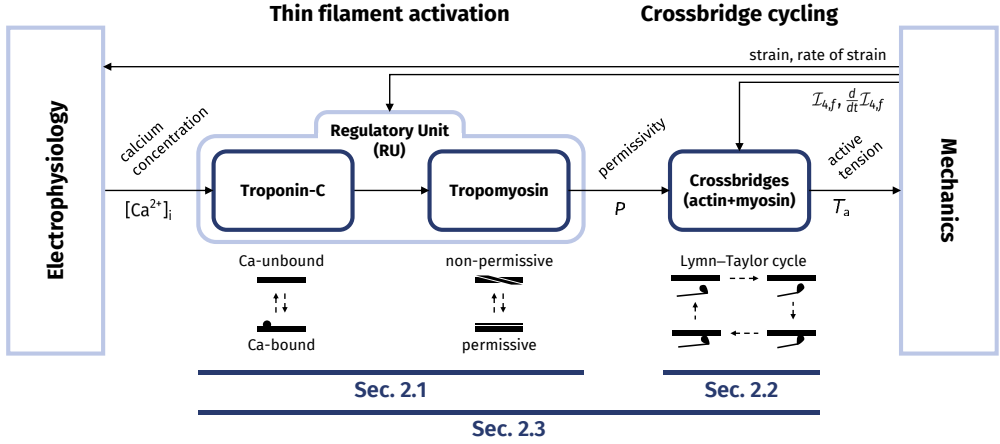

Figure 1: Representation of the different stages of the force generation mechanism and of the sections where they are discussed in this paper.

The construction of an integrated mathematical and numerical model of cardiac electromechanics (EM) is however a remarkably arduous task. This is mainly due to the *multiphysics* (due to the interplay of biochemistry, electricity, solid mechanics, fluid dynamics) and *multiscale* nature of the heart: characteristic spatial scales range from nanometers to centimeters and the temporal ones from microseconds to seconds. This makes it difficult to devise computationally efficient and accurate algorithms for a plurality of mathematical models featuring a broad degree of details [3, 16, 26, 35, 79, 93].

The contrasting needs between model accuracy and computational efficiency of numerical simulations is mainly due to the multiscale nature of the heart, for which the mechanical work enabling the macroscopic motion of the organ is fueled by the energy consumed at the microscale by subcellular mechanisms. The generation of active force takes place inside sarcomeres and involves a complex chain of chemical and mechanical reactions. This mechanism can be split into two steps, that we sketch in Fig. 1. First, a ionic signal (specifically, an increase of calcium ions concentration) triggers the so-called *regulatory units*, protein complexes consisting of troponin and tropomyosin, that act as on-off switches for the muscle contraction. Then, when the regulatory units are activated, the *actin* and *myosin* proteins are free to interact and form the so-called *crossbridges*, molecular motors that generate an active force by consuming the chemical energy stored in ATP [7, 52].

Microscopic force generation includes many regulatory mechanisms, forming the subcellular basis of organ-level phenomena, such as the Frank-Starling effect [52]. Hence, if a microscale mathematical model of force generation is used in a multiscale setting to build an integrated organ-level EM model, then it should be able to reproduce the above-mentioned mechanisms.

In the past decades, several efforts have been dedicated to the construction of mathematical models describing the complex dynamics of the processes taking place in sarcomeres [13–15, 17, 44, 46, 47, 61, 62, 66, 67, 86, 88, 90, 108, 109]. How-ever, because of the intrinsic complexity of the phenomenon of force generation, huge computational costs are associated with the numerical approximation of such models, thus limiting their application within multiscale EM simulations. Despite several attempts to capture the fundamental mechanisms underlying the force generation phenomenon into a tractable number of equations [11, 60, 80, 88, 90, 92, 107], the existing organ-level cardiac mathematical models rely on two alternative strategies to describe microscopic force generation.

- Phenomenological models (see e.g. [43, 61, 62, 67, 89]) are built by fitting the measured data with simple laws, chosen by the modeler. However, the parameters characterizing phenomenological models often lack a clear physical interpretation; moreover, the noisy nature and deficiency of data coming from the subcellular contractile units and the intrinsic difficulties in measuring sarcomeres under the conditions occurring during an heartbeat hamper the predictive power of such models [31].

- Biophysically detailed models are based on an accurate description of the proteins involved in the force generation process and are derived from physics first principles. However, their numerical solution, because of their complexity, is typically obtained by means of a Monte Carlo (MC) approximation (see e.g. [44, 108, 109]). The MC method is in fact inefficient, featuring a huge computational cost, both in terms of time and memory storage. Indeed, to accurately approximate the solution of a single heartbeat for a single myofilament, tens of hours of computational time may be required; see e.g. [81].

The purpose of this paper is to develop a biophysically detailed model for active force generation, that explicitly describes the fundamental ingredients of the force generation apparatus, yet with a tractable computational cost, so that it is suitable for multiscale EM simulations.

## 1.1 Paper outline

This paper is structured as follows. In Sec. 2 we recall the main features of the force generation phenomenon in cardiomyocytes and the main difficulties encountered in the construction of mathematical models describing the associated mechanisms. In Sec. 3 we present the models proposed in this paper and in Sec. 4 we describe the strategy employed for their calibration. Then, in Sec. 5, we show some numerical results obtained with the proposed models, including filament-level 0D simulations and multiscale 3D cardiac EM simulations. We provide some final remarks in Sec. 6. In Tab. 1 we provide a list of the abbreviations used throughout this paper.

## 2 Microscale models of cardiac contraction

We recall the microscopical mechanisms by which active force is generated in the cardiac tissue and we highlight the difficulties, rooted in the their intrinsic complexity, in describing such phenomena with a tractable number of equations. We also review the main contributions available in literature.

Sarcomeres are cylindrically-shaped, 2 µm length units made of a regular arrangement of thick and thin filaments. The former, also known as myosin filaments (MF), are mainly made of the protein myosin, while the latter are made of actin, troponin (Tn) and tropomyosin (Tm) and are also called actin filaments (AF).

| Anatomical terms | | Mathematical models | |
|---|---|---|---|
| LV | Left ventricle | H57 | Model of [47] |
| Tm | Tropomyosin | TTP06 | Model of [99] |
| Tn | Troponin | ToR-ORd | Model of [104] |
| MH | Myosin head | | |
| XB | Crossbridge | **Others abbreviations** | |
| RU | Regulatory unit | ODE | Ordinary differential equation |
| BS | Binding site | PDE | Partial differential equation |
| MF | Myosin (thick) filament | CTMC | Continuous-time Markov Chain |
| AF | Actin (thin) filament | MC | Monte Carlo |
| EM | Electromechanics | | |

Table 1: List of abbreviations of this paper.

In Sec. 2.1 we deal with the activation of the thin filament, involving the troponin-tropomyosin regulatory units (RUs). Then, in Sec. 2.2, we address the crossbridge (XB) cycling and finally, in Sec. 2.3, we consider the full-sarcomere dynamics (see Fig. 1).

## 2.1 Modeling the thin filament activation

The activation of the thin filament is mainly regulated by two variables, namely the intracellular calcium ions concentration ($[\mathrm{Ca}^{2+}]_\mathrm{i}$) and the sarcomere length ($SL$). The experimental signature of the regulation mechanisms is given by the steady-steady relationships between calcium, sarcomere length and generated force.

The force-calcium relationship (see Fig. 2) features a sigmoidal shape, well described by the Hill equation [27, 55, 102]:

$$T_\mathrm{a}^\mathrm{iso} = \frac{T_\mathrm{a}^\mathrm{max}}{1 + \left(\frac{\mathrm{EC}_{50}}{[\mathrm{Ca}^{2+}]_\mathrm{i}}\right)^{n_H}}, \qquad (1)$$

where $T_\mathrm{a}^\mathrm{max}$ is the maximum tension (tension at saturating calcium levels), $\mathrm{EC}_{50}$ is the half maximal effective concentration (i.e. the calcium concentration producing half maximal force) and $n_H$ is the Hill coefficient. The experimentally measured force-calcium curves in the cardiac tissue feature a steep slope in correspondence of half activation (Hill coefficient greater than one) [4, 27, 102], thus revealing the presence of cooperative effects. Despite several explanations have been proposed [24, 29, 30, 34, 38, 87, 88, 98], the most likely hypothesis lies in the end-to-end interactions of Tm units [40, 71, 94].

An increase of $SL$ has a two-fold effect on the steady-state tension (see Fig. 2): the plateau force (i.e. the tension associated with saturating calcium) increases and the calcium-sensitivity is enhanced (i.e. the sigmoidal curves are left-ward shifted). Whereas the explanation of the first effect is commonly-agreed to be linked to the increase of extension of the single-overlap zone (i.e. the region of the sarcomere where the MF filament a single AF), a well-assessed explanation for the second effect (known as length-dependent activation, LDA) has not yet been found [1, 10, 30, 33, 68, 72, 87, 94, 100, 103].

The earliest attempts to model the calcium-driven regulation of the muscular contractile system date back to the 1990s [28, 63, 80, 88, 92, 112]. Those models rely on the formalism of continuous-time Markov Chains (CTMC), also known as Markov Jump processes (see e.g. [69]), to model the transitions between the different configurations assumed by the proteins involved in the force regulation process. In those models, the necessity of representing the end-to-end interactions of Tm units dictates a spatially-explicit representation of the RUs. Indeed, mean-field models, where only a single representative RU is considered, fail to correctly predict the cooperative activation and the resulting steep force-calcium curves [87, 90], unless phenomenological laws are introduced, as discussed below. However, the number of degrees of freedom of the CTMC increases exponentially with the number of RUs represented. This hinders the possibility of numerically approximating the solution of the Forward Kolmogorov Equation, also known as Master Equation in natural sciences, which describes the time evolution of the probabilities associated with the states of a stochastic process [5]. As a matter of fact, the Forward Kolmogorov Equation associated with this CTMC would have a number of variables that is exponential in the number of RUs, resulting in as many as $10^{20}$ or more variables [81]. For this reason, spatially-explicit models require a MC approximation for their numerical resolution, thus resulting in very large computational costs [81, 87, 107].

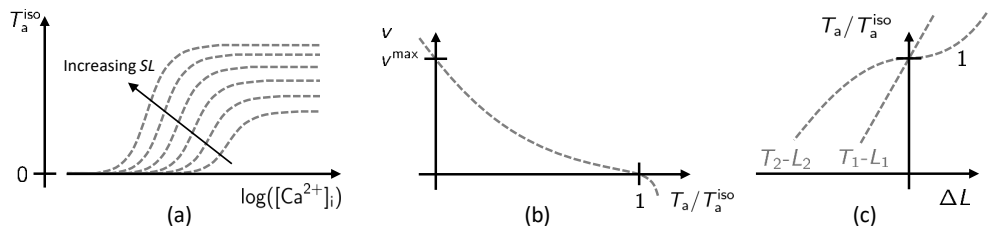

Figure 2: Representation of the steady-state force-calcium relationship (a), the force-velocity relationship (b) and tension-elongation curves after a fast transient (c).

To avoid an explicit representation of end-to-end interactions, phenomenological models, where the transition rates are set as a nonlinear functions of the calcium concentration, have been proposed [43, 61, 62, 67, 89]. These models are however based on phenomenological laws. Alternatively, to overcome the large computational cost induced by the MC method without renouncing to represent end-to-end interactions (unlike in phenomenological models), several attempts to capture cooperative phenomena by means of numerically tractable ODE systems have been done in literature. In [90] an analytical solution is derived for the steady-state. In [11], a periodicity assumption is used to reduce the number of unknowns. In [107] each RU is considered independently of each others, while end-to-end interactions are accounted for by fitting the parameters of an integro-differential system with memory from a collection of simulations. In [60], the states of the CTMC are grouped by the number of unblocked RUs and a MC sampling technique is used to estimate the average free energy of each group and, thus, the transition rates within groups. For further details on these modeling attempts, the interested reader can refer to [81, 85].

## 2.2 Modeling the crossbridge dynamics

Active force is generated by XBs by the cyclical attachment and detachment of myosin heads (MHs) to actin binding sites (BSs). When MHs are in their attached configuration, they rotate towards the center of the sarcomere, performing the so-called power-stroke, thus pulling the AF along the same direction. Such cyclical path, known as Lymn-Taylor cycle [64], features a wide range of time scales (nearly from 1 to 100 ms) [7, 15, 53]; hence, also the response of the force generation apparatus to external stimuli is characterized by different time scales. Indeed, when a fast step in force (respectively, in length) is applied to an isometrically contracted muscle fiber, three different phases can be observed [12, 15, 53, 65, 66]. First, an instantaneous elastic response occurs along the so-called $T_1$-$L_1$ curve (see Fig. 2), whose slope corresponds to the stiffness of the attached myosin proteins. Then, a fast transient $(2-3\,\mathrm{ms})$ occurs, and the sarcomere reaches a length (respectively, a tension) belonging to the $T_2$-$L_2$ curve (see Fig. 2). Such second phase corresponds to a rearrangement of MHs within their pre- and post-power-stroke configuration. Finally, with a time scale on nearly 100 ms, the muscle fiber reaches a steady-state regime, characterized (in case the step is applied by controlling the force) by a constant shortening (or lengthening) velocity. The relationship between the steady-state velocity and the muscle tension constitutes the so-called force-velocity curve (see Fig. 2), firstly measured by Hill in [41], and it is characterized by a finite value of velocity (denoted by $v^{\mathrm{max}}$) for which the generated tension is zero [7, 52]. The experimental measurements of the $T_1$-$L_1$, the $T_2$-$L_2$ and the force-velocity curves are invariant after normalization of the tension by its isometric value, denoted by $T_{\mathrm{a}}^{\mathrm{iso}}$. This reveals that the underlying mechanisms are related to the XB dynamics, while they are independent of the thin filament regulation, whose effect is simply that of tuning the number of recruitable XBs [7, 12, 52].

The attachment-detachment process of MHs has been described accordingly with the formalism of the Huxley model [47] (that we denote by H57 model), where the population of MHs is described by the distribution density of the distortion of attached XB. The time evolution of such distribution is driven by a PDE, where a convection term accounts for the mutual sliding between filaments, and a source and a sink term (whose rates depend on the XB distortion) account for the creation and destruction of XBs [17, 45, 58, 59]. In order to capture the separation between the fastest time scales (i.e. between the first two phases following a fast step either in force or in length), an explicit representation of the power-stroke must be included in the model, by introducing a multistable discrete [46, 95] or continuous [14, 15, 57, 65, 66] degree of freedom, representing the angular position of the rotating MH.

### 2.3 Modeling the full sarcomere dynamics

In the past two decades, several models describing the generation of active force in the cardiac tissue, including both the calcium-driven regulation and the XB cycling, have been proposed. The main challenge faced in the development of such models lies in the spatial dependence of the cooperativity phenomenon, crucial to reproduce the calcium dependence of muscle activation (see Sec. 2.1). As a matter of fact, an explicit representation of spatial-dependent cooperative mechanisms dramatically increases the computational complexity of activation models, even more so when such models are coupled with models describing XB cycling. When the interactions between BSs and MHs are considered, indeed, one must face the further difficulty of tracking which BS faces which MH when the filaments mutually slide. The attempt of capturing such spatially dependent phenomena in a compact system of ODEs is the common thread of most of the literature on sarcomere modeling (see e.g. [8, 11, 17, 60, 81, 89–91, 107, 111]). We remark that computational efficiency is a major issue when sarcomere models are employed in multiscale simulation, such as cardiac EM. In this case, indeed, the microscale force generation model needs to be simultaneously solved in many points of the computational domain (typically at each nodal point of the computational mesh). Nonetheless, most of biophysically detailed full-sarcomere models rely on the time-consuming MC method for their numerical approximation [44, 97, 107–109].

## 3 Proposed full-sarcomere models

In this section, we propose four different microscale models of active force generation in the cardiac tissue. These models are derived from a biophysically detailed CTMC, accurately describing the dynamics of the regulatory and contractile proteins. We present a strategy to derive a set of equations, with a dramatically smaller number of variables than the Forward Kolmogorov Equation, describing the evolution of the biophysically detailed CTMC. Our strategy is based on physically motivated assumptions, aimed at neglecting second-order interactions among the proteins, focusing only on the main interactions, so that the variables describing the stochastic processes associated with the states of the proteins can be partially decoupled, similarly to what we have done in [81]. This results in a drastic reduction in the size of models. Moreover, we change the classical MH-centered description of XBs (see e.g. [14, 45, 46, 95]), in favor of a BS-centered one. This prevents the necessity of tracking the mutual sliding between the filaments, still without the need of introducing any simplifying assumption.

As in most of RUs models (see Sec. 2.1), we describe Tn and Tm by discrete states. Moreover, based on the experimental evidence that cooperativity is linked to RUs end-to-end interactions [40, 71, 94], we include nearest-neighbor interactions among RUs with the formalism of the model of [90].

Concerning the modeling of XBs, we are here interested in developing a model of cardiomyocytes contraction in the heart, which is characterized by slower time-scales than the fast time-scale of the power-stroke. This suggests that the level of detail that best suits the application to cardiac EM does not require to explicitly represent the power-stroke [82]. In [14], indeed, the authors showed that, if the time-scales of interest are slower than the time-scale of the power-stroke, the detailed models including the power-stroke reduce to H57-like models, where only the attachment-detachment process of XBs is explicitly represented. Therefore, we model the XB dynamics as a two-states process, within the H57 formalism, where the attachment-detachment rates depend on the myosin arm distortion.

### 3.1 Model setup

We consider a pair of interacting myofilaments and we denote by $N_A$ the number of RUs located on an AF and by $N_M$ the number of MHs located on half MF. To identify a RU we employ the index $i \in \mathcal{I}_A := \{1, \dots, N_A\}$, while to identify MHs we employ the index $j \in \mathcal{I}_M := \{1, \dots, N_M\}$. The CTMC model describing the dynamics of the RUs and of the MHs is sketched in Fig. 3.

Each RU is composed by a Tn unit and by a Tm unit, respectively associated with the variables $C_i^t$ and $T_i^t$. Tn can be either unbound ($\mathcal{U}$) or bound ($\mathcal{B}$) to calcium (we write $C_i^t = \mathcal{U}$ and $C_i^t = \mathcal{B}$, respectively). On the other hand, Tm can be either in non-permissive ($\mathcal{N}$) or in permissive ($\mathcal{P}$) configuration (we write $T_i^t = \mathcal{N}$ and $T_i^t = \mathcal{P}$, respectively). In our model, the calcium binding kinetics is affected by the state of Tm. Hence, when the $i$-th Tm unit is non-permissive (i.e. $T_i^t = \mathcal{N}$), we denote the binding and unbinding calcium rates by $k_{C,i}^{\mathcal{UB}|\mathcal{N}}$ and $k_{C,i}^{\mathcal{BU}|\mathcal{N}}$, respectively; conversely, when $T_i^t = \mathcal{P}$, we denote the binding and unbinding calcium rates by $k_{C,i}^{\mathcal{UB}|\mathcal{P}}$ and $k_{C,i}^{\mathcal{BU}|\mathcal{P}}$. Similarly, the kinetics of Tm is affected by the calcium-binding state of the corresponding Tn unit. Moreover, because of the Tm end-to-end interactions, the Tm transition rates also depend on the state of the nearest-neighboring Tm units. Hence, the transition rates from the non-permissive to the permissive states and vice versa – given the state of $T_{i-1}$, $T_{i+1}$ and $C_i$ – are respectively denoted by $k_{T,i}^{\mathcal{NP}|T_{i-1}\cdot T_{i+1}, C_i}$ and $k_{T,i}^{\mathcal{PN}|T_{i-1}\cdot T_{i+1}, C_i}$. To better clarify the notation, an example is shown in Fig. 3 (bottom-left box). In the example, $k_{T,i}^{\mathcal{NP}|\mathcal{P}\cdot\mathcal{P},\mathcal{B}}$ denotes the transition rate for $i$-th Tm unit from the non-permissive to the permissive state, when the nearest-neighboring units are both permissive and the associated Tn unit is bound to calcium. Concerning the Tm units located at the ends of the filaments, for which a neighbor is missing, the latter is assumed to be in state $\mathcal{N}$. We exclude any feedback from XBs on the dynamics of the RUs, as recent experimental evidence suggests that this kind of feedback is not present [30, 98].

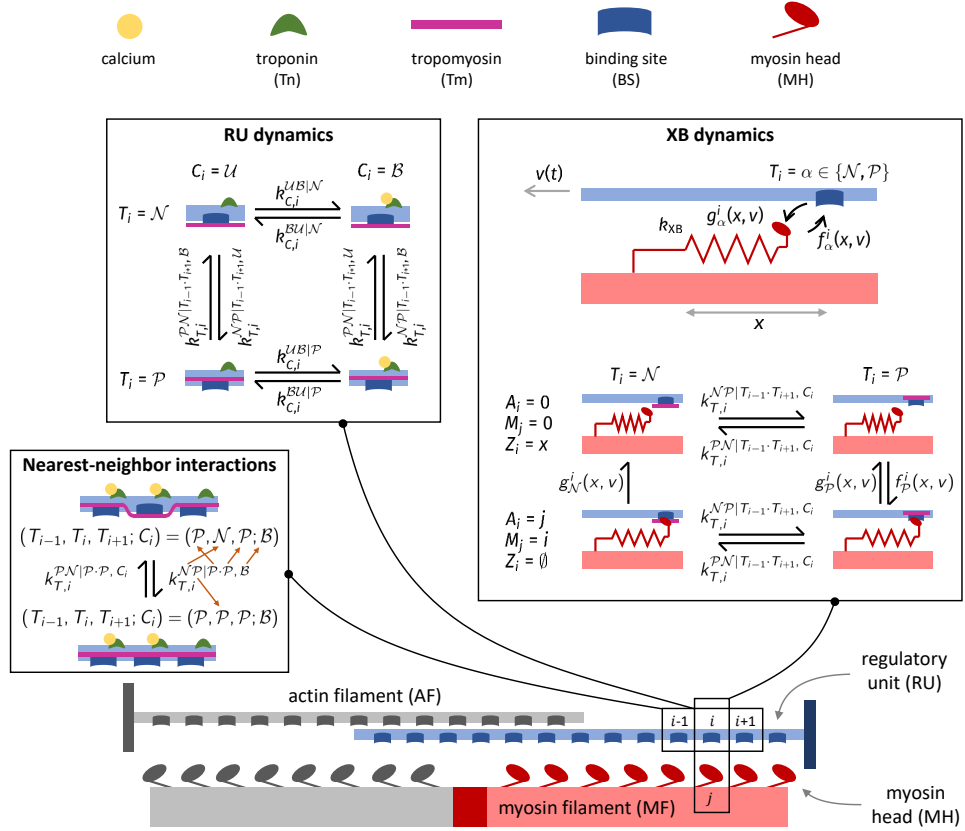

Figure 3: Sketch of the proposed CTMC model. Each RU is described by a 4-state model (top left), whose dynamics is affected by the state nearest-neighboring RUs. An example of nearest-neighbor interactions is shown in the bottom-left box, where the notation for the transition rates $k_{T,i}$ is illustrated by the orange arrows. MHs are described as 2-state elements, whose transition rates are affected by the XB elongation, the sliding velocity between myofilaments and the permissivity state of the RU.

Each myosin arm is modeled as a linear spring with stiffness $k_{\text{XB}}$. The attachment and detachment rates of XBs depend on the distance between the MH resting position and the BS, denoted by $x$, and on the relative velocity between the myofilaments, denoted by $v_{\text{hs}}(t) := -\frac{d}{dt}SL(t)/2$. For simplicity, we define as $v(t) = 2\, v_{\text{hs}}(t)/SL_0$ the normalized shortening velocity, where $SL_0$ denotes a reference sarcomere length. Moreover, to model the calcium-driven regulation, the XB attachment and detachment rates depend on the state of the corresponding Tm unit. Hence, when $T_i^t = \mathcal{N}$, we denote the XB binding and unbinding rates by $f_{\mathcal{N}}^i(x, v(t))$ and $g_{\mathcal{N}}^i(x, v(t))$, respectively; conversely, when $T_i^t = \mathcal{P}$, we denote the XB binding and unbinding rates by $f_{\mathcal{P}}^i(x, v(t))$ and $g_{\mathcal{P}}^i(x, v(t))$. In particular, we set $f_{\mathcal{N}}^i \equiv 0$ since new XBs can form only if the corresponding Tm unit is permissive. Clearly, a XB can form only if neither the BS nor the MH is already attached to another site. Moreover, the attachment rates $f_{\mathcal{N}}^i$ and $f_{\mathcal{P}}^i$ are zero sufficiently far from $x = 0$.

In order to describe the state of XBs, we introduce the variables $A_i^t$, $M_j^t$ and $Z_i^t$, respectively denoting the state of actin BSs, the state of MHs and the displacement of attached XBs. Specifically, when the $i$-th actin BS is attached to the $j$-th MH we write $A_i^t = j$. Similarly, when the $j$-th MH is attached to the $i$-th actin BS we write $M_j^t = i$. Moreover, we write $Z_i^t = x$ whenever the $i$-th actin BS is attached to a MH with displacement $x$. Clearly, these variables are redundant, as we have:

$$(A_i^t = j) \iff (M_j^t = i) \iff (Z_i^t = d_{ij}(t)),$$

where we have denoted by $d_{ij}(t)$ the distance between the $i$-th BS and the $j$-th MH.

To summarize, the CTMC is described by the following stochastic processes, for

$i \in \mathcal{I}_A$, $j \in \mathcal{I}_M$ and $t \geq 0$:

$$C_i^t = \begin{cases} \mathcal{B} & \text{if the } i\text{-th Tn is bound to calcium,} \\ \mathcal{U} & \text{otherwise;} \end{cases}$$

$$T_i^t = \begin{cases} \mathcal{P} & \text{if the } i\text{-th Tm is permissive,} \\ \mathcal{N} & \text{otherwise;} \end{cases}$$

$$A_i^t = \begin{cases} j & \text{if the } i\text{-th actin BS is attached to the } j\text{-th MH,} \\ 0 & \text{if the } i\text{-th actin BS is not attached to any MH;} \end{cases} \quad (2)$$

$$M_j^t = \begin{cases} i & \text{if the } j\text{-th MH is attached to the } i\text{-th actin BS,} \\ 0 & \text{if the } j\text{-th MH is not attached to any actin BS;}, \end{cases}$$

$$Z_i^t = \begin{cases} x & \text{if the } i\text{-th actin BS is attached to a MH with displacement } x, \\ \emptyset & \text{if the } i\text{-th actin BS is not attached to any MH.} \end{cases}$$

We remark that we denote the detached state by $Z_i^t = \emptyset$ rather than $Z_i^t = 0$, because the latter notation is employed to denote the case when the $i$-th actin BS is attached with displacement $x = 0$.

The total force exerted by the pair of interacting half MF and AF is given by the sum of the force generated by each attached XB. Therefore, the expected value of the force is given by:

$$F_{\text{hf}}(t) = \sum_{i \in \mathcal{I}_A} \mathbb{E}\left[F_{\text{XB}}(Z_i^t)\right],$$

where $F_{\text{XB}}(x)$ denotes the force exerted by an attached MH with distortion $x$ and where we set by convention $F_{\text{XB}}(\emptyset) = 0$. Here and in what follows, we denote by $\mathbb{E}[\cdot]$ the expected value of a random variable.

The size of the CTMC (2), that is the number of its states, is overwhelming. As a matter of fact, each RU can be in four possible states ($\mathcal{UN}$, $\mathcal{BN}$, $\mathcal{UP}$ and $\mathcal{BP}$), and the corresponding BS can be either unbound or bound to one of the $N_M$ MHs. In conclusion, the number of states of the CTMC is $(4(N_M + 1))^{N_A} \simeq 10^{60}$. Thus, the numerical solution of the associated Forward Kolmogorov Equation is unaffordable, as it would feature as many variables as the number of states of the CTCM [81]. To overcome this inconvenient, we introduce some physical motivated assumptions that allow to partially decouple the dynamics of the stochastic processes, thus yielding a much reduced set of equations.

## 3.2 Models equations

In this section, we present equations describing the evolution of the stochastic processes of Eq. (2). A detailed derivation of these equations is provided in App. A.

### 3.2.1 Thin filament regulation

Due to the lack of feedback from XBs to RUs, it is possible to write an equation describing the evolution of the stochastic processes $C_i^t$ and $T_i^t$ independently of the stochastic processes associated with the XBs (while the converse clearly does not hold). Similarly to [81], we focus on the joint probabilities of triplets of consecutive RUs.

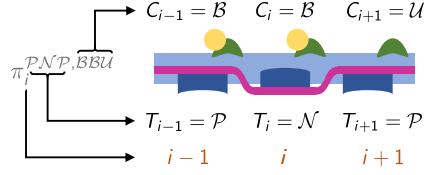

Figure 4: Representation of the configuration corresponding to the state variable $\pi_i^{\mathcal{PNP},\mathcal{BBU}}(t)$. The arrows illustrate the meaning of the notation.

Hence, we consider the following variables, where $i = 2,\ldots,N_A - 1$, $\vartheta,\eta,\lambda \in \{\mathcal{U},\mathcal{B}\}$ and $\alpha,\beta,\delta \in \{\mathcal{N},\mathcal{P}\}$:

$$\pi_i^{\alpha\beta\delta,\vartheta\eta\lambda}(t) := \mathbb{P}\left[(T_{i-1},T_i,T_{i+1})^t = (\alpha,\beta,\delta), (C_{i-1},C_i,C_{i+1})^t = (\vartheta,\eta,\lambda)\right], \qquad (3)$$

where $\mathbb{P}[\cdot]$ denotes the probability of an event. For instance, $\pi_i^{\mathcal{PNP},\mathcal{BBU}}(t)$ denotes the probability that the triplet centered in the $i$-th unit has the Tm units in states $\mathcal{P}$, $\mathcal{N}$ and $\mathcal{P}$ and the Tn units in states $\mathcal{B}$, $\mathcal{B}$ and $\mathcal{U}$ respectively, as shown in Fig. 4.

For each $i = 2,\ldots,N_A - 1$, we have 64 variables written in the form (3), corresponding to as many states of the triplet. The dynamics of each variable is determined by 6 possible forward and backward transitions, as depicted in Fig. 5. However, the transition rates associated with the Tm units at the edge of the triplet cannot be computed from the variables of Eq. (3), as they depend on the state of a Tm unit outside the triplet. For instance, the rate of the transition from $\pi_i^{\mathcal{PNP},\mathcal{BBU}}(t)$ to $\pi_i^{\mathcal{PNN},\mathcal{BBU}}(t)$ depends on the state of $T_{i+2}^t$, which does not belong to the triplet. Nonetheless, under a suitable hypothesis, the transition rates between Tm being in permissive and non-permissive state can be defined as:

$$\widetilde{k}_{T,i}^{\overline{\alpha}\alpha|\circ\cdot\beta,\circ\vartheta\eta} := \begin{cases} \dfrac{\sum_{\xi,\zeta} k_{T,i}^{\overline{\alpha}\alpha|\xi\cdot\beta,\vartheta} \pi_i^{\xi\overline{\alpha}\beta,\zeta\vartheta\eta}}{\sum_{\xi,\zeta} \pi_i^{\xi\overline{\alpha}\beta,\zeta\vartheta\eta}} & \text{for } i = 2,\ldots,N_A - 1, \\[1em] k_{T,i}^{\overline{\alpha}\alpha|\mathcal{N}\cdot\beta,\vartheta} & \text{for } i = 1; \end{cases}$$

$$\widetilde{k}_{T,i}^{\overline{\delta}\delta|\beta\cdot\circ,\eta\lambda\circ} := \begin{cases} \dfrac{\sum_{\xi,\zeta} k_{T,i}^{\overline{\delta}\delta|\beta\cdot\xi,\lambda} \pi_i^{\beta\overline{\delta}\xi,\eta\lambda\zeta}}{\sum_{\xi,\zeta} \pi_i^{\beta\overline{\delta}\xi,\eta\lambda\zeta}} & \text{for } i = 2,\ldots,N_A - 1, \\[1em] k_{T,i}^{\overline{\delta}\delta|\beta\cdot\mathcal{N},\lambda} & \text{for } i = N_A, \end{cases} \qquad (4)$$

where the symbol $\circ$ recalls that the corresponding unit has an arbitrary state. In Eq. (4) and in what follows, we use the notation $\overline{\mathcal{N}} = \mathcal{P}$, $\overline{\mathcal{P}} = \mathcal{N}$, $\overline{\mathcal{U}} = \mathcal{B}$ and $\overline{\mathcal{B}} = \mathcal{U}$ to denote opposite states. For instance, if $\alpha = \mathcal{N}$, then $\overline{\alpha} = \mathcal{P}$.

Equation (4) is rigorously derived in the Supporting Information (S1 Appendix). The derivation is rather technical and is based upon the following assumption:

$$\begin{aligned} (T_{i+1},C_{i+1})^t \perp\!\!\!\perp T_{i-2}^t | (T_{i-1},T_i,C_{i-1},C_i)^t & \quad \text{for } i = 3,\ldots,N_A - 1, \\ (T_{i-1},C_{i-1})^t \perp\!\!\!\perp T_{i+2}^t | (T_{i+1},T_i,C_{i+1},C_i)^t & \quad \text{for } i = 2,\ldots,N_A - 2, \end{aligned} \qquad \text{(H1)}$$

where, given three events $A,B,C$ we write $A \perp\!\!\!\perp B|C$ if $A$ and $B$ are conditionally independent given $C$ (i.e. $\mathbb{P}[A \cap B|C] = \mathbb{P}[A|C]\mathbb{P}[B|C]$). Assumption (H1) states that distant RUs are conditionally independent given the states of the intermediate RUs. From a modeling viewpoint, this means that the interaction of far units is mediated

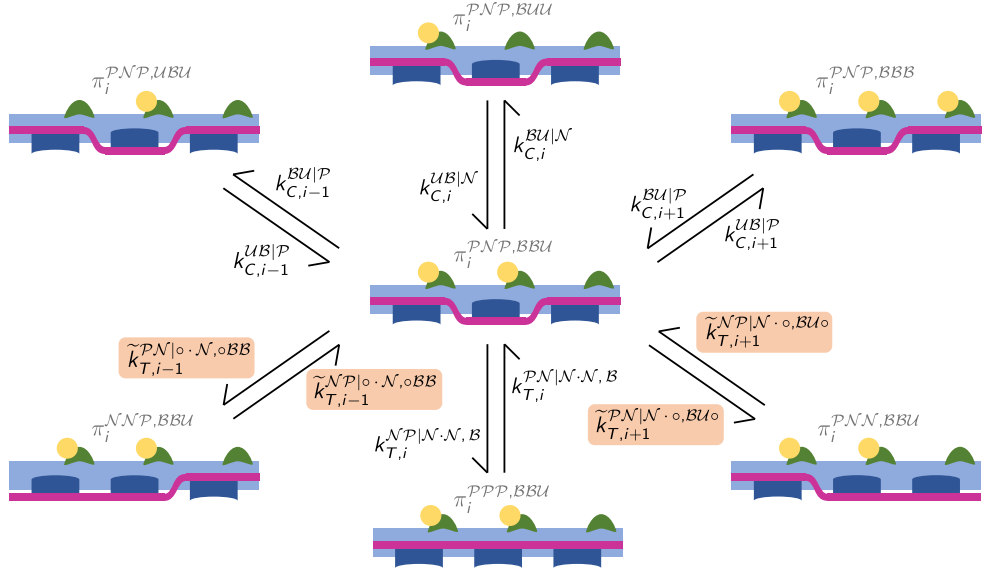

Figure 5: Spatially-explicit model: each triplet of consecutive RUs can undergo 6 different transitions. For example, this figure shows the transitions of the configuration associated with the state variable $\pi_i^{\mathcal{PNP},\mathcal{BBU}}(t)$, with the corresponding transition rates. The transition rates computed thanks to Ass. (H1) are highlighted with a colored background.

by the intermediate ones, which is coherent with the short-range nature of end-to-end interactions.

In conclusion, we obtain the following system of ODEs, for $t \geq 0$ and for any $i = 2, \ldots, N_A - 1$, $\vartheta, \eta, \lambda \in \{\mathcal{U}, \mathcal{B}\}$ and $\alpha, \beta, \delta \in \{\mathcal{N}, \mathcal{P}\}$:

$$\begin{aligned}
\frac{d}{dt}\pi_i^{\alpha\beta\delta,\vartheta\eta\lambda} =\ & \widetilde{k}_{T,i-1}^{\overline{\alpha}\alpha|\circ\,\cdot\,\beta,\circ\vartheta\eta} && \pi_i^{\overline{\alpha}\beta\delta,\vartheta\eta\lambda} && - \widetilde{k}_{T,i-1}^{\alpha\overline{\alpha}|\circ\,\cdot\,\beta,\circ\vartheta\eta} && \pi_i^{\alpha\beta\delta,\vartheta\eta\lambda} \\
& + k_{T,i}^{\overline{\beta}\beta|\alpha\,\cdot\,\delta,\eta} && \pi_i^{\alpha\overline{\beta}\delta,\vartheta\eta\lambda} && - k_{T,i}^{\beta\overline{\beta}|\alpha\,\cdot\,\delta,\eta} && \pi_i^{\alpha\beta\delta,\vartheta\eta\lambda} \\
& + \widetilde{k}_{T,i+1}^{\overline{\delta}\delta|\beta\,\cdot\,\circ,\eta\lambda\circ} && \pi_i^{\alpha\beta\overline{\delta},\vartheta\eta\lambda} && - \widetilde{k}_{T,i+1}^{\delta\overline{\delta}|\beta\,\cdot\,\circ,\eta\lambda\circ} && \pi_i^{\alpha\beta\delta,\vartheta\eta\lambda} \\
& + k_{C,i-1}^{\overline{\vartheta}\vartheta|\alpha} && \pi_i^{\alpha\beta\delta,\overline{\vartheta}\eta\lambda} && - k_{C,i-1}^{\vartheta\overline{\vartheta}|\alpha} && \pi_i^{\alpha\beta\delta,\vartheta\eta\lambda} \\
& + k_{C,i}^{\overline{\eta}\eta|\beta} && \pi_i^{\alpha\beta\delta,\vartheta\overline{\eta}\lambda} && - k_{C,i}^{\eta\overline{\eta}|\beta} && \pi_i^{\alpha\beta\delta,\vartheta\eta\lambda} \\
& + k_{C,i+1}^{\overline{\lambda}\lambda|\delta} && \pi_i^{\alpha\beta\delta,\vartheta\eta\overline{\lambda}} && - k_{C,i+1}^{\lambda\overline{\lambda}|\delta} && \pi_i^{\alpha\beta\delta,\vartheta\eta\lambda},
\end{aligned} \quad (5)$$

endowed with suitable initial conditions.

The permissivity of a given RU is defined as its probability of being in permissive state (i.e. $P_i(t) = \mathbb{P}\left[T_i^t = \mathcal{P}\right]$) and can be obtained from the variables $\pi_i^{\alpha\beta\delta,\vartheta\eta\lambda}(t)$ as:

$$P_i(t) = \begin{cases} \displaystyle\sum_{\beta,\delta\in\{\mathcal{N},\mathcal{P}\}} \sum_{\vartheta,\eta,\lambda\in\{\mathcal{U},\mathcal{B}\}} \pi_2^{\mathcal{P}\beta\delta,\vartheta\eta\lambda}(t) & \text{for } i = 1, \\ \displaystyle\sum_{\alpha,\delta\in\{\mathcal{N},\mathcal{P}\}} \sum_{\vartheta,\eta,\lambda\in\{\mathcal{U},\mathcal{B}\}} \pi_i^{\alpha\mathcal{P}\delta,\vartheta\eta\lambda}(t) & \text{for } i = 2,\ldots,N_A - 1, \\ \displaystyle\sum_{\alpha,\beta\in\{\mathcal{N},\mathcal{P}\}} \sum_{\vartheta,\eta,\lambda\in\{\mathcal{U},\mathcal{B}\}} \pi_{N_A-1}^{\alpha\beta\mathcal{P},\vartheta\eta\lambda}(t) & \text{for } i = N_A. \end{cases}$$

### 3.2.2 Crossbridge dynamics

Similarly to the H57 model, we introduce distribution density functions tracking the elongation of attached XBs. However, since in our CTMC the XB transition rates depend on the state of the associates Tm unit, we split attached XBs into two families: those associated with a non-permissive Tm and those associated with a permissive one. Hence, we define the following variables, for $i \in \mathcal{I}_A$, corresponding to the probability density that the $i$-th BS is attached to a MH with displacement $x$ and that the associated RU is in a given permissivity state:

$$\begin{aligned} n_{i,\mathcal{P}}(x,t) &= \mathbb{f}\left[Z_i^t = x, T_i^t = \mathcal{P}\right], \\ n_{i,\mathcal{N}}(x,t) &= \mathbb{f}\left[Z_i^t = x, T_i^t = \mathcal{N}\right], \end{aligned} \tag{6}$$

where the symbol $\mathbb{f}[\cdot]$ denotes a probability density function.

We notice that we make here the choice of tracking the XBs from the point of view of the BSs, rather than of the MHs, as it is traditionally done in literature [14, 45, 46, 95]. This change of perspective has the significant advantage that it does not require to track which RU faces which MH at each time. Indeed, each BS and each RU, being located on the same filament, rigidly move with respect of each others and, thus, each BS is always associated with the same RU.

In App. A.4 we derive the following system of PDEs, describing the time evolution of $n_{i,\mathcal{P}}$ and $n_{i,\mathcal{P}}$:

$$\begin{cases} \dfrac{\partial n_{i,\mathcal{P}}}{\partial t} - v_{\text{hs}}\dfrac{\partial n_{i,\mathcal{P}}}{\partial x} = (D_M^{-1}P_i - n_{i,\mathcal{P}})f_\mathcal{P}^i - g_\mathcal{P}^i n_{i,\mathcal{P}} \\ \qquad\qquad - \widetilde{k}_{T,i}^{\mathcal{P}\mathcal{N}} n_{i,\mathcal{P}} + \widetilde{k}_{T,i}^{\mathcal{N}\mathcal{P}} n_{i,\mathcal{N}} & x \in \mathbb{R},\ t \geq 0,\ i \in \mathcal{I}_A, \\ \dfrac{\partial n_{i,\mathcal{N}}}{\partial t} - v_{\text{hs}}\dfrac{\partial n_{i,\mathcal{N}}}{\partial x} = (D_M^{-1}(1 - P_i) - n_{i,\mathcal{N}})f_\mathcal{N}^i - g_\mathcal{N}^i n_{i,\mathcal{N}} \\ \qquad\qquad - \widetilde{k}_{T,i}^{\mathcal{N}\mathcal{P}} n_{i,\mathcal{N}} + \widetilde{k}_{T,i}^{\mathcal{P}\mathcal{N}} n_{i,\mathcal{P}} & x \in \mathbb{R},\ t \geq 0,\ i \in \mathcal{I}_A, \end{cases} \tag{7}$$

endowed with suitable initial conditions, where we define:

$$\begin{aligned} \widetilde{k}_{T,i}^{\mathcal{N}\mathcal{P}} &:= \frac{\sum_{\alpha,\delta,\vartheta,\eta,\lambda} k_{T,i}^{\mathcal{N}\mathcal{P}|\alpha\cdot\delta,\eta} \pi_i^{\alpha\mathcal{N}\delta,\vartheta\eta\lambda}}{1 - P_i}, \\ \widetilde{k}_{T,i}^{\mathcal{P}\mathcal{N}} &:= \frac{\sum_{\alpha,\delta,\vartheta,\eta,\lambda} k_{T,i}^{\mathcal{P}\mathcal{N}|\alpha\cdot\delta,\eta} \pi_i^{\alpha\mathcal{P}\delta,\vartheta\eta\lambda}}{P_i}. \end{aligned} \tag{8}$$

The sink and source terms in Eq. (7) account for the fluxes among the two groups (XBs with non-permissive Tm and with permissive Tm). The terms $P_i$ and $(1 - P_i)$ represent the maximum possible proportion of attached BSs in each group. The term $D_M$, defined as the distance between two consecutive MHs, appears because in this setting $n_{i,\mathcal{P}}$ and $n_{i,\mathcal{N}}$ are, from a dimensional point of view, the inverse of length units (they are probability densities), whereas the variables of the H57 model are dimensionless. We remark that, differently than the H57 model, Eq. (7) is referred to BSs rather than to MHs.

The derivation of Eq. (7), presented in App. A.4, is based on two assumptions. First, we assume that the state of a BS is conditionally independent of the state of surrounding RUs, given the permissivity state of the associated RU. This is coherent with the physics of the model, as the only feature of the RUs that directly affects the XBs binding rates is the permissivity state of Tm. In mathematical terms, this assumption reads:

$$A_i^t \perp\!\!\!\perp (T_{i-1}, T_{i+1}, C_i)^t | T_i^t \qquad \text{for } i = 2, \ldots, N_A - 1. \tag{H2}$$

Second, we assume that the shortening velocity $v$ is never so large to convect attached BSs within the range of attachment of a different MH. In mathematical terms, we assume that:

$$f_{\mathcal{P}}^i(d_{ij}(t), v(t)) \neq 0 \implies A_h \neq j \quad \forall h \neq i. \tag{H3}$$

The latter assumption allows to decouple the dynamics of the different units. We notice that all the models belonging to the family of the H57 model are based on assumptions analogous to Ass. (H3), without which the H57 equation cannot be derived.

By combining Eq. (5) with Eq. (7), describing the dynamics of RUs and XBs, respectively, we obtain a model that we denote as the *SE-PDE model* (where *SE* stands for spatially-explicit, while *PDE* denotes the fact that the XB dynamics is described by a PDE system). In this model, the expected value of the force exerted by the whole half filament can be determined as follows:

$$F_{\text{hf}}(t) = \sum_{i \in \mathcal{I}_A} \int_{-\infty}^{+\infty} F_{\text{XB}}(x) \left(n_{i,\mathcal{P}}(x,t) + n_{i,\mathcal{N}}(x,t)\right) dx. \tag{9}$$

### 3.2.3 Distribution-moments equations

When the XB attachment-detachment transition rates assume special forms, the PDE system of Eq. (7) can be reduced to a more compact system of ODEs, by following a general strategy in statistical physics, already used for H57-like models [8, 17, 111]. Specifically, under suitable hypotheses on the transition rates, the distributions of the elongation of attached XBs (i.e. $n_{i,\mathcal{P}}(x,t)$ and $n_{i,\mathcal{N}}(x,t)$) can be fully characterized by their first two moments that we denote by $\mu_{i,\mathcal{P}}^0$, $\mu_{i,\mathcal{P}}^1$ and by $\mu_{i,\mathcal{N}}^0$, $\mu_{i,\mathcal{N}}^1$, respectively. More precisely, we define, for $\alpha \in \{\mathcal{N}, \mathcal{P}\}$, for $\psi \in \{f_{\mathcal{P}}^i, f_{\mathcal{N}}^i, g_{\mathcal{P}}^i, g_{\mathcal{N}}^i\}$ and for $p = 0, 1$:

$$\begin{aligned}
\mu_{i,\alpha}^p(t) &:= \int_{-\infty}^{+\infty} \left(\frac{x}{SL_0/2}\right)^p n_{i,\alpha}(x,t) dx, \\
\mu_{\psi}^p(v) &:= \int_{-\infty}^{+\infty} \left(\frac{x}{SL_0/2}\right)^p \psi(x,v) \frac{dx}{D_M}.
\end{aligned} \tag{10}$$

We notice that the zero order moment $\mu_{i,\mathcal{N}}^0(t)$ (respectively, $\mu_{i,\mathcal{P}}^0(t)$) can be interpreted as the probability that the $i$-th BS is attached and associated to a non-permissive (respectively, permissive) RU. Moreover, the ratio $\mu_{i,\mathcal{N}}^1(t)/\mu_{i,\mathcal{N}}^0(t)$ (respectively, $\mu_{i,\mathcal{P}}^1(t)/\mu_{i,\mathcal{P}}^0(t)$) corresponds to the expected value of the distortion (normalized with respect to $SL_0/2$ of the XB attached to the $i$-th RU, provided that the corresponding RU is in non-permissive (respectively, permissive) state.

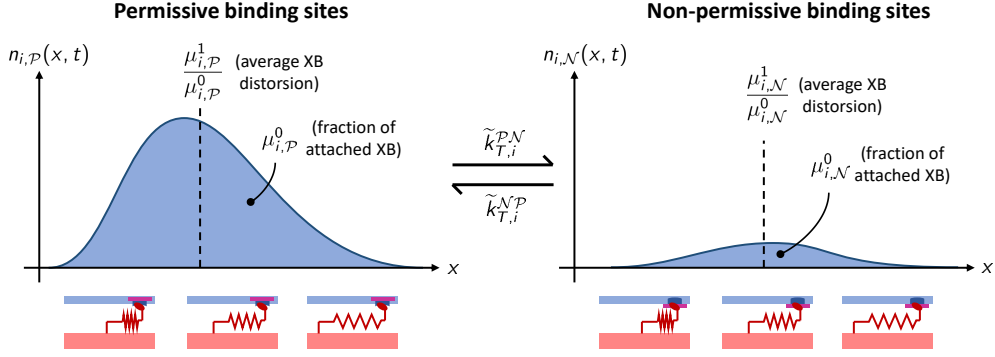

Figure 6: Representation of the variables of the distribution-moments equations. The variable $\mu^0_{i,\mathcal{P}}(t)$ (respectively, $\mu^0_{i,\mathcal{N}}(t)$) corresponds to the fraction of permissive (respectively, non-permissive) BSs to which a MH is bound, while the ratio $\mu^1_{i,\mathcal{P}}(t)/\mu^0_{i,\mathcal{P}}(t)$ (respectively, $\mu^1_{i,\mathcal{N}}(t)/\mu^0_{i,\mathcal{N}}(t)$) corresponds to the average XB distortion within the permissive (respectively, non-permissive) attached BSs.

Let us assume that the total attachment-detachment rate is independent of the XB distortion (i.e. there exist functions $r^i_{\mathcal{P}}(v)$ and $r^i_{\mathcal{N}}(v)$, for $i \in \mathcal{I}_A$, such that $r^i_{\mathcal{P}}(v) = f^i_{\mathcal{P}}(x,v) + g^i_{\mathcal{P}}(x,v)$ and $r^i_{\mathcal{N}}(v) = f^i_{\mathcal{N}}(x,v) + g^i_{\mathcal{N}}(x,v)$ for any $x \in \mathbb{R}$). Under this assumption, as shown in App. A.5, we get the following distribution-moments equations:

$$\begin{cases} \dfrac{d}{dt}\mu^0_{i,\mathcal{P}} = -\left(r^i_{\mathcal{P}} + \widetilde{k}^{\mathcal{PN}}_{T,i}\right)\mu^0_{i,\mathcal{P}} + \widetilde{k}^{\mathcal{NP}}_{T,i}\mu^0_{i,\mathcal{N}} + P_i \mu^0_{f^i_{\mathcal{P}}} & t \geq 0,\ i \in \mathcal{I}_A, \\ \dfrac{d}{dt}\mu^0_{i,\mathcal{N}} = -\left(r^i_{\mathcal{N}} + \widetilde{k}^{\mathcal{NP}}_{T,i}\right)\mu^0_{i,\mathcal{N}} + \widetilde{k}^{\mathcal{PN}}_{T,i}\mu^0_{i,\mathcal{P}} + (1-P_i) \mu^0_{f^i_{\mathcal{N}}} & t \geq 0,\ i \in \mathcal{I}_A, \\ \dfrac{d}{dt}\mu^1_{i,\mathcal{P}} + v\,\mu^0_{i,\mathcal{P}} = -\left(r^i_{\mathcal{P}} + \widetilde{k}^{\mathcal{PN}}_{T,i}\right)\mu^1_{i,\mathcal{P}} + \widetilde{k}^{\mathcal{NP}}_{T,i}\mu^1_{i,\mathcal{N}} + P_i \mu^1_{f^i_{\mathcal{P}}} & t \geq 0,\ i \in \mathcal{I}_A, \\ \dfrac{d}{dt}\mu^1_{i,\mathcal{N}} + v\,\mu^0_{i,\mathcal{N}} = -\left(r^i_{\mathcal{N}} + \widetilde{k}^{\mathcal{NP}}_{T,i}\right)\mu^1_{i,\mathcal{N}} + \widetilde{k}^{\mathcal{PN}}_{T,i}\mu^1_{i,\mathcal{P}} + (1-P_i) \mu^1_{f^i_{\mathcal{N}}} & t \geq 0,\ i \in \mathcal{I}_A. \end{cases} \quad (11)$$

By assuming a linear spring hypothesis for the tension generated by attached XBs (i.e. $F_{\text{XB}}(x) = k_{\text{XB}}\,x$), the expected value of the force of half filament is given by:

$$F_{\text{hf}}(t) = k_{\text{XB}} \frac{SL_0}{2} N_A \mu^1(t), \quad (12)$$

where we have defined:

$$\mu^p(t) = \frac{1}{N_A}\sum_{i=1}^{N_A}\left(\mu^p_{i,\mathcal{P}} + \mu^p_{i,\mathcal{N}}\right). \quad (13)$$

Since the active force generated by half MF is proportional to $\mu^1(t)$, there exists some constant $a_{\text{XB}}$ (with the dimension of a pressure) such that the macroscopic active tension can be written as $T_a(t) = a_{\text{XB}}\,\mu^1(t)$. In the following, we denote by *SE-ODE model* the one obtained by combining Eq. (5) with Eq. (11) (where *ODE* denotes the fact that the XB dynamics is described by an ODE system).

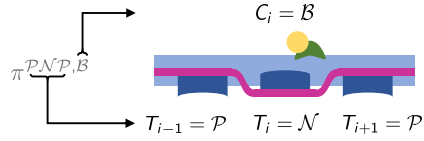

Figure 7: Representation of the configuration corresponding to the state variable $\pi^{\mathcal{P}\mathcal{N}\mathcal{P},\mathcal{B}}(t)$. The arrows illustrate the meaning of the notation.

### 3.2.4 Mean-field approximation

The models proposed so far are based on an explicit spatial description of the physical arrangements of proteins along the myofilaments. The spatial description allows to model the cooperativity mechanism (linked to the nearest-neighbor interactions within RUs) and the $SL$ related effects on the force generation machinery (linked to the filament overlapping). However, the first phenomenon, despite being spatially dependent, is based on interactions of local type; the effect of the second phenomenon, in turn, largely depends on the size of the single-overlap zone, that is a scalar quantity non dependent on the spatial variable. Based on the above considerations, we propose a mean-field approximation of the spatially dependent CTMC presented in Sec. 3.1, where the nearest-neighbor interaction are captured as a local effect, and the effect of $SL$ is modeled in function of the size of the single-overlap zone.

This mean-field model is based on the assumption that the single-overlap zone can be considered as infinitely long. Such approximation is reasonable as far as the effect of the edges can be neglected (the validity of such approximation will be discussed in Sec. 6). A direct consequence of this assumption is the invariance by translation of the joint distribution of RUs. In other terms, the variables $\pi_i^{\alpha\beta\delta,\vartheta\eta\lambda}(t)$ defined in Eq. (3) coincide for each $i$. In addition, we further reduce the number of variables by tracking only the state of the Tn unit of the central RU of the triplet (this further reduction is made possible by the fact that we never have to track the behavior at the boundaries of the filaments, as we will see in what follows). We define thus the following variables, for $\alpha, \beta, \delta \in \{\mathcal{N}, \mathcal{P}\}$ and $\eta \in \{\mathcal{U}, \mathcal{B}\}$:

$$\pi^{\alpha\beta\delta,\eta}(t) := \mathbb{P}\left[(T_{i-1}, T_i, T_{i+1})^t = (\alpha, \beta, \delta), C_i^t = \eta\right]. \tag{14}$$

A visual representation of these variables is provided in Fig. 7. We notice that the variables $\pi^{\alpha\beta\delta,\eta}(t)$ are well-defined thanks to the translational invariance of the distribution of RUs. Moreover, the transition rates $k_{C,i}^{\delta\bar{\delta}|\beta}$ and $k_{T,i}^{\beta\bar{\beta}|\alpha\cdot\eta,\delta}$ for the units in the single-overlap region do not depend on $i$. Hence, we will denote them simply as $k_C^{\delta\bar{\delta}|\beta}$ and $k_T^{\beta\bar{\beta}|\alpha\cdot\eta,\delta}$.

The dynamics of the variables $\pi^{\alpha\beta\delta,\eta}(t)$ involves 4 possible forward and backward transitions (see Fig. 8). Similarly to the spatially-explicit model, the transitions rates associated with the edge Tm units cannot be determined without additional assumptions. Therefore, we introduce the following assumption:

$$\begin{aligned}
(T_{i+1}, C_i)^t &\perp\!\!\!\perp (T_{i-2}, C_{i-1})^t | (T_{i-1}, T_i)^t && \text{for } i = 3, \ldots, N_A - 1, \\
(T_{i-1}, C_i)^t &\perp\!\!\!\perp (T_{i+2}, C_{i+1})^t | (T_{i+1}, T_i)^t && \text{for } i = 2, \ldots, N_A - 2,
\end{aligned} \tag{H4}$$

Assumption (H4), similarly to (H1), states the conditional independence of far units, given the states of the intermediate ones. In other terms, by Assumption (H4) we neglect the long-range interactions between Tm units, which is a secondary effect compared to short-range (i.e. end-to-end) interactions. We remark that Ass. (H4) and (H1) are two different mathematical translations of the same physical concept. The difference is due to the different state representation done in the two models: while the state of the SE model comprises three Tn units, the one of the MF model comprises only one Tn unit.

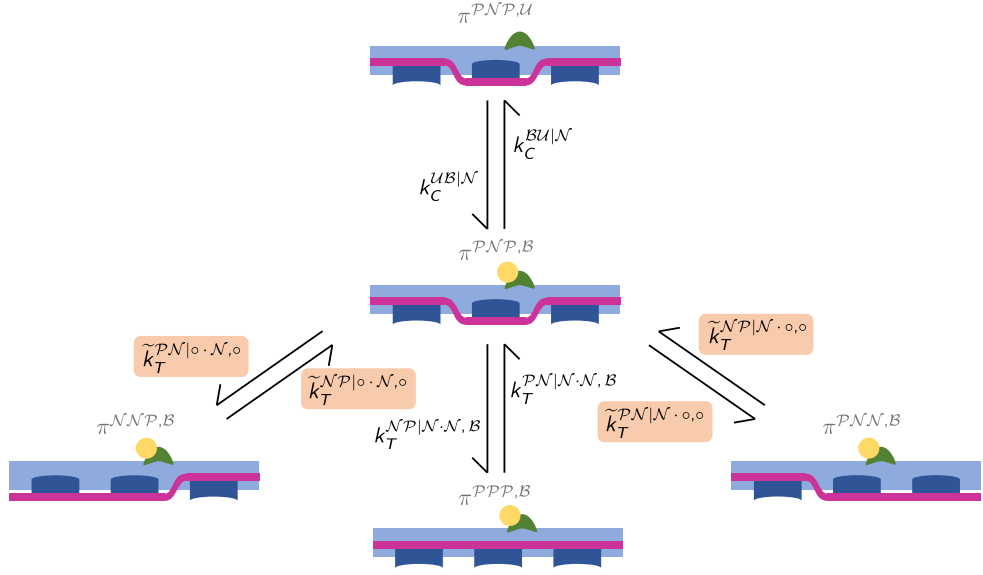

Figure 8: Mean-field model: representation of the 4 forward and backward transitions of the configuration associated with the state variable $\pi^{\mathcal{PNP},\mathcal{B}}(t)$, with the corresponding transition rates. The transition rates computed thanks to Ass. (H4) are highlighted with a colored background.

In this way we obtain (as proved in App. A.3), the following ODE model, valid for $t \geq 0$ and for any $\alpha, \beta, \delta \in \{\mathcal{N}, \mathcal{P}\}$ and $\eta \in \{\mathcal{U}, \mathcal{B}\}$:

$$\begin{aligned}
\frac{d}{dt}\pi^{\alpha\beta\delta,\eta} &= \widetilde{k}_T^{\overline{\alpha}\alpha|\circ\cdot\beta,\circ}\,\pi^{\overline{\alpha}\beta\delta,\eta} - \widetilde{k}_T^{\alpha\overline{\alpha}|\circ\cdot\beta,\circ}\,\pi^{\alpha\beta\delta,\eta} \\
&+ k_T^{\overline{\beta}\beta|\alpha\cdot\delta,\eta}\,\pi^{\alpha\overline{\beta}\delta,\eta} - k_T^{\beta\overline{\beta}|\alpha\cdot\delta,\eta}\,\pi^{\alpha\beta\delta,\eta} \\
&+ \widetilde{k}_T^{\overline{\delta}\delta|\beta\cdot\circ,\circ}\,\pi^{\alpha\beta\overline{\delta},\eta} - \widetilde{k}_T^{\delta\overline{\delta}|\beta\cdot\circ,\circ}\,\pi^{\alpha\beta\delta,\eta} \\
&+ k_C^{\overline{\eta}\eta|\beta}\,\pi^{\alpha\beta\delta,\overline{\eta}} - k_C^{\eta\overline{\eta}|\beta}\,\pi^{\alpha\beta\delta,\eta},
\end{aligned} \qquad (15)$$

where:

$$\widetilde{k}_T^{\overline{\alpha}\alpha|\circ\cdot\beta,\circ} := \frac{\sum_{\xi,\zeta} k_T^{\overline{\alpha}\alpha|\xi\cdot\beta,\zeta}\pi^{\xi\overline{\alpha}\beta,\zeta}}{\sum_{\xi,\zeta}\pi^{\xi\overline{\alpha}\beta,\zeta}},$$

$$\widetilde{k}_T^{\overline{\delta}\delta|\beta\cdot\circ,\circ} := \frac{\sum_{\xi,\zeta} k_T^{\overline{\delta}\delta|\beta\cdot\xi,\zeta}\pi^{\beta\overline{\delta}\xi,\zeta}}{\sum_{\xi,\zeta}\pi^{\beta\overline{\delta}\xi,\zeta}}.$$

The permissivity of a RU in the single-overlap zone, defined as $P(t) = \mathbb{P}[T_i^t = \mathcal{P}]$

(such that the $i$-th RU belongs to the single-overlap zone), can be obtained as:

$$P(t) = \sum_{\alpha,\delta \in \{\mathcal{N},\mathcal{P}\}} \sum_{\eta \in \{\mathcal{U},\mathcal{B}\}} \pi^{\alpha\mathcal{P}\delta,\eta}(t).$$

By similar arguments, it follows that also the joint distribution of the stochastic processes associated with XB formation enjoys the translational invariance property and, consequently, the following variables are well defined, as the right-hand sides are independent of the index $i$ (for $i$ belonging to the single-overlap zone):

$$\begin{aligned} n_{\mathcal{P}}(x,t) &= \mathbb{f}\left[Z_i^t = x, T_i^t = \mathcal{P}\right], \\ n_{\mathcal{N}}(x,t) &= \mathbb{f}\left[Z_i^t = x, T_i^t = \mathcal{N}\right]. \end{aligned} \quad (16)$$

By proceeding as in Sec. 3.2.2, we get the following model:

$$\begin{cases} \dfrac{\partial n_{\mathcal{P}}}{\partial t} - v_{\text{hs}} \dfrac{\partial n_{\mathcal{P}}}{\partial x} = (D_M{}^{-1}P - n_{\mathcal{P}})f_{\mathcal{P}} - g_{\mathcal{P}} n_{\mathcal{P}} \\ \qquad\qquad - \widetilde{k}_T^{\mathcal{PN}} n_{\mathcal{P}} + \widetilde{k}_T^{\mathcal{NP}} n_{\mathcal{N}} & x \in \mathbb{R},\, t \geq 0, \\ \dfrac{\partial n_{\mathcal{N}}}{\partial t} - v_{\text{hs}} \dfrac{\partial n_{\mathcal{N}}}{\partial x} = (D_M{}^{-1}(1-P) - n_{\mathcal{N}})f_{\mathcal{N}} - g_{\mathcal{N}} n_{\mathcal{N}} \\ \qquad\qquad - \widetilde{k}_T^{\mathcal{NP}} n_{\mathcal{N}} + \widetilde{k}_T^{\mathcal{PN}} n_{\mathcal{P}} & x \in \mathbb{R},\, t \geq 0, \end{cases} \quad (17)$$

where we have defined:

$$\begin{aligned} \widetilde{k}_T^{\mathcal{NP}}(t) &:= \dfrac{\sum_{\alpha,\delta,\eta} k_T^{\mathcal{NP}|\alpha\cdot\delta,\eta}\, \pi^{\alpha\mathcal{N}\delta,\eta}(t)}{1 - P(t)}, \\ \widetilde{k}_T^{\mathcal{PN}}(t) &:= \dfrac{\sum_{\alpha,\delta,\eta} k_T^{\mathcal{PN}|\alpha\cdot\delta,\eta}\, \pi^{\alpha\mathcal{P}\delta,\eta}(t)}{P(t)}. \end{aligned} \quad (18)$$

The expected value of the force exerted by the whole half filament can be obtained as follows:

$$F_{\text{hf}}(t) = N_A\, \chi_{\text{so}}(SL(t)) \int_{-\infty}^{+\infty} F_{\text{XB}}(x)\, (n_{\mathcal{P}}(x,t) + n_{\mathcal{N}}(x,t))\, dx, \quad (19)$$

where the single-overlap ratio $\chi_{\text{so}}$ denotes the fraction of the AF filament in the single-overlap zone:

$$\chi_{\text{so}}(SL) := \begin{cases} 0 & \text{if } SL \leq L_A, \\ \dfrac{2(SL - L_A)}{L_M - L_H} & \text{if } L_A < SL \leq L_M, \\ \dfrac{SL + L_M - 2L_A}{L_M - L_H} & \text{if } L_M < SL \leq 2L_A - L_H, \\ 1 & \text{if } 2L_A - L_H < SL \leq 2L_A + L_H, \\ \dfrac{L_M + 2L_A - SL}{L_M - L_H} & \text{if } 2L_A + L_H < SL \leq 2L_A + L_M, \\ 0 & \text{if } SL > 2L_A + L_M, \end{cases} \quad (20)$$

$L_A$ being the length of the AF, $L_M$ the length of the MF and $L_H$ the length of the bare zone (see Fig. 9). We notice that we are here assuming that the relative sliding between the filaments is such that $\chi_{\mathrm{so}}$ slowly varies, so that we can neglect the effects linked to the state transitions taking place at the border of the single-overlap zone. The combination of Eqs. (15) and (17) gives a model for the full-sarcomere dynamics, which we denote as the *MF-PDE model* (where *MF* stands for mean-field).

Moreover, under the assumption that the total attachment-detachment rate does not depend on the XB elongation (i.e. there exist two functions $r_\mathcal{P}(v)$ and $r_\mathcal{N}(v)$ such that $r_\mathcal{P}(v) = f_\mathcal{P}(x,v) + g_\mathcal{P}(x,v)$ and $r_\mathcal{N}(v) = f_\mathcal{N}(x,v) + g_\mathcal{N}(x,v)$ for any $x \in \mathbb{R}$), we can derive the following distribution-moment equation:

$$\begin{cases} \dfrac{d}{dt}\mu_\mathcal{P}^0 = -\left(r_\mathcal{P}(v) + \widetilde{k}_T^{\mathcal{PN}}\right)\mu_\mathcal{P}^0 + \widetilde{k}_T^{\mathcal{NP}}\mu_\mathcal{N}^0 + P\,\mu_{f_\mathcal{P}}^0 & t \geq 0, \\[6pt] \dfrac{d}{dt}\mu_\mathcal{N}^0 = -\left(r_\mathcal{N}(v) + \widetilde{k}_T^{\mathcal{NP}}\right)\mu_\mathcal{N}^0 + \widetilde{k}_T^{\mathcal{PN}}\mu_\mathcal{P}^0 + (1-P)\,\mu_{f_\mathcal{N}}^0 & t \geq 0, \\[6pt] \dfrac{d}{dt}\mu_\mathcal{P}^1 + v\,\mu_\mathcal{P}^0 = -\left(r_\mathcal{P}(v) + \widetilde{k}_T^{\mathcal{PN}}\right)\mu_\mathcal{P}^1 + \widetilde{k}_T^{\mathcal{NP}}\mu_\mathcal{N}^1 + P\,\mu_{f_\mathcal{P}}^1 & t \geq 0, \\[6pt] \dfrac{d}{dt}\mu_\mathcal{N}^1 + v\,\mu_\mathcal{N}^0 = -\left(r_\mathcal{N}(v) + \widetilde{k}_T^{\mathcal{NP}}\right)\mu_\mathcal{N}^1 + \widetilde{k}_T^{\mathcal{PN}}\mu_\mathcal{P}^1 + (1-P)\,\mu_{f_\mathcal{N}}^1 & t \geq 0, \end{cases} \quad (21)$$

endowed with suitable initial conditions and where we define, for $\alpha \in \{\mathcal{N}, \mathcal{P}\}$:

$$\mu_\alpha^p(t) := \int_{-\infty}^{+\infty} \left(\frac{x}{SL_0/2}\right)^p n_\alpha(x,t)\,dx. \qquad (22)$$

The force exerted by half thick filament is then given by:

$$F_{\mathrm{hf}}(t) = k_{\mathrm{XB}} \frac{SL_0}{2} N_A\, \mu^1(t), \qquad (23)$$

where

$$\mu^p(t) := \chi_{\mathrm{so}}(SL(t))\left[\mu_\mathcal{P}^p(t) + \mu_\mathcal{N}^p(t)\right], \qquad (24)$$

for $p = 0, 1$ and, hence, the tissue level active tension is $T_\mathrm{a}(t) = a_{\mathrm{XB}}\,\mu^1(t)$. Finally, by combining Eqs. (15) and (21) we obtain a model that we denote as *MF-ODE model*.

### 3.2.5 List of proposed models

Table 2 provides a recap of the different models proposed in this paper. For each model, we report the assumptions underlying its derivation. The two models derived within the distribution-moments formalism (SE-ODE and MF-ODE) also require that the sum of the attachment and detachment rates is independent of $x$ (we write $f + g \perp\!\!\!\perp x$). We notice that this is not a simplificatory assumption, but rather a specific modeling choice.

Table 2 contains four different models describing the same biological phenomenon. A natural question is when each model is to be preferred with respect to the others and which are its advantages and disadvantages. The modeler should make two choices:

- *ODE versus PDE models.* The two PDE models leave a great freedom to the modeler in the choice of the XB transition rate functions $f_\mathcal{P}(x,v)$, $f_\mathcal{N}(x,v)$, $g_\mathcal{P}(x,v)$ and $g_\mathcal{N}(x,v)$. Hence, if the modeler has knowledge of the specific form of these functions (or if he wants to test different choices), PDE models are to be preferred. Otherwise, the two ODE models allow to avoid having to define the precise form of the functions $f_\mathcal{P}(x,v)$, $f_\mathcal{N}(x,v)$, $g_\mathcal{P}(x,v)$ and $g_\mathcal{N}(x,v)$, only requiring to set a few scalar parameters, that can be easily calibrated from macroscopic measurements, as we show later.

| Model name (Equations) | Number of ODEs Number of PDEs | Assumptions | Modeling choices |
|---|---|---|---|
| **SE-PDE** (5)-(7) | $(N_A - 2)2^6 = 1280$ $2N_A = 64$ | (H1),(H2),(H3) | |
| **SE-ODE** (5)-(11) | $(N_A - 2)2^6 + 4N_A = 1408$ - | (H1),(H2),(H3) | $f + g \perp\!\!\!\perp x$ |
| **MF-PDE** (15)-(17) | $2^4 = 16$ $2 = 2$ | (H4),(H2),(H3), m.f. | |
| **MF-ODE** (15)-(21) | $2^4 + 4 = 20$ - | (H4),(H2),(H3), m.f. | $f + g \perp\!\!\!\perp x$ |

Table 2: List of the models proposed in Sec. 3. For future reference, we assign a name to each model (*SE* stands for spatially-explicit, *MF* stands for mean-field). In the second column we report the number of ODEs and PDEs included in each model as a function of $N_A$ and $N_M$ and we specify the resulting values in the case $N_A = 32$, $N_M = 18$. In the "Assumptions" column, *m.f.* stands for *mean-field assumption*.

> If the interest of the modeler is oriented towards the microscopical dynamics of XBs, PDE models allow to simulate the precise distribution of XB strains, whereas ODE models only describe its first two moments. The greater detail of PDE models, however, comes at the price of larger computational costs. Therefore, in the context of multiscale simulations such as cardiac electromechanics – where the model should be solved simultaneously in a large number of points – the ODE models are computationally more attractive than PDE models.

- *SE (spatially-explicit) versus MF (mean-field) models.* The two MF models are derived from the corresponding SE models by considering a single triplet of RUs rather than a whole AF. Hence, the former models neglect the profile that the states of RUs assume alongside the AF. As we show later, this profile – specifically the behavior near the end-points of the single-overlap zone – may play a role in the *SL*-induced change in calcium sensitivity (the so-called LDA), but the sources of this phenomenon have not been not fully understood yet [1, 68, 94, 103]. Clearly, the MF models are much lighter than the SE ones. Hence, the former should be preferred when computational cost is a major issue, such as in large-scale simulations of cardiac electromechanics. On the other hand, SE models are to be preferred if the modeler is interested in studying the activation profile alongside the AFs (e.g. to investigate its role in the onset of the LDA phenomenon).

### 3.3 Definition of transition rates

In Sec. 3.2 we have shown that, under some physically motivated assumptions, the CTMC presented in Sec. 3.1 can be described by different systems of ODEs and/or PDEs. The models listed in Tab. 2 are thus valid independently of the specific choice of the transition rates (with the only exception of the models SE-ODE and MF-ODE that require that the sum of the detachment and attachment rate is independent of the XB distorsion). In this section, we present and motivate the specific choice of transition rates that we will adopt in the rest of this paper.

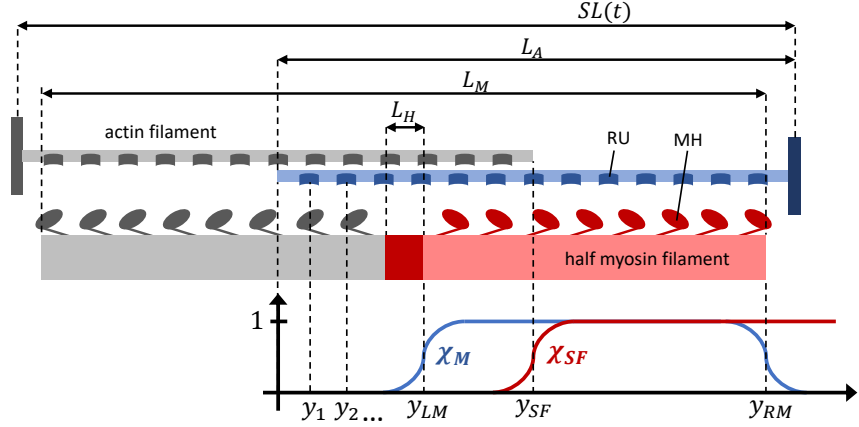

Figure 9: Sketch of the sarcomere model. The thick filament (MF) is represented in red and two thin filaments (AF) are represented in blue (top). The origin of the frame of reference is the left side of the reference AF. The functions $\chi_{SF}$ and $\chi_M$ are also represented (bottom).

In the spatially-explicit models introduced in Sec. 3.1, the transition rates possibly depend on the location of the RUs in the myofilaments. This allows to account for the myofilaments overlap (single overlap, duble overlap, no overlap). To facilitate the identification of the overlap region corresponding to a given RU from its index $i \in \mathcal{I}_A$, we introduce the following functions, which define a smooth transition between the regions (see Fig. 9):

$$\chi_M(SL, i) = \frac{1}{2} \tanh\left(\frac{y_i - y_{LM}}{\varepsilon}\right) + \frac{1}{2} \tanh\left(-\frac{y_i - y_{RM}}{\varepsilon}\right),$$
$$\chi_{SF}(SL, i) = \frac{1}{2}\left(1 + \tanh\left(\frac{y_i - y_{SF}}{\varepsilon}\right)\right),$$

where we have defined:

$$y_{LM} = (2 L_A - SL + L_H)/2, \qquad y_{SF} = 2 L_A - SL,$$
$$y_{RM} = (2 L_A - SL + L_M)/2, \qquad y_i = \frac{L_A}{N_A}(i - 0.5).$$

as the coordinates (with respect to the end of the AF closer to the center of the sarcomere) of the left and right ends of the MF ($y_{LM}$, $y_{RM}$), of the beginning of the single-overlap region ($y_{SF}$) and of the $i$-th RU ($y_i$). Hence, we have $\chi_M(SL, i) \simeq 1$ if the $i$-th RU faces the considered half MF and $\chi_{SF}(SL, i) \simeq 1$ if the $i$-th RU is in the single filament region (no overlap with other AFs occurs).

### 3.3.1 RUs transition rates

The RU dynamics is determined by the eight rates associated with the forward and backward transitions $\mathcal{UN} \rightleftharpoons \mathcal{BN}$, $\mathcal{UP} \rightleftharpoons \mathcal{BP}$, $\mathcal{UN} \rightleftharpoons \mathcal{UP}$ and $\mathcal{BN} \rightleftharpoons \mathcal{BP}$. The transition rates are affected by $[Ca^{2+}]_i$ (that enhances in a multiplicative way the transition $\mathcal{U} \rightarrow \mathcal{B}$), the filament overlap and the state of the nearest-neighboring Tm units (for the latter interaction we adopt the cooperative interactions proposed in [90]). We start by considering the single-overlap zone, where we adopt the transition rates of the model of [90]. We show below that the transition rates of [90] are, however, rather general, as they are based on just a couple of assumptions. We keep the notation consistent with [90] to allow for comparisons.

We call $k_C^{\mathcal{BU}|\mathcal{N}} := k_{\text{off}}$ and, without loss of generality, we set $k_C^{\mathcal{BU}|\mathcal{P}} := k_{\text{off}}/\mu$, where the constant $\mu$ allows to differentiate the two rates. Experiments carried out with protein isoforms from different species highlight that there is no apparent variation in the transition $\mathcal{U} \to \mathcal{B}$ in different combinations of Tn subunits and Tm [67]. We assume thus that the transition rates for $\mathcal{UN} \to \mathcal{BN}$ and for $\mathcal{UP} \to \mathcal{BP}$ coincide, and we set $k_C^{\mathcal{UB}|\mathcal{N}} = k_C^{\mathcal{UB}|\mathcal{P}} := k_{\text{off}}/k_{\text{d}} [\text{Ca}^{2+}]_{\text{i}}$. Conversely, we allow the reverse transition rates to depend on the state of the associated Tm. Concerning the transitions involving Tm, we assume that the calcium binding state of Tn affects the transition rate of $\mathcal{N} \to \mathcal{P}$ for the associated Tm, but not the reverse rate. Therefore, we set $k_T^{\mathcal{PN}|\alpha\cdot\delta,\mathcal{U}} = k_T^{\mathcal{PN}|\alpha\cdot\delta,\mathcal{B}} = k_{\text{basic}}\gamma^{2-n(\alpha,\delta)}$, where $n(\alpha,\delta) \in \{0,1,2\}$ denotes the number of permissive states among $\alpha$ and $\delta$, as proposed in [90]. Then, without loss of generality we denote $k_T^{\mathcal{NP}|\alpha\cdot\delta,\mathcal{B}} = Q\,k_{\text{basic}}\gamma^{n(\alpha,\delta)}$, where the constant $Q$ allows to differentiate the forward and backward transition rates*. The only transition rate left is given, to satisfy the detailed-balance consistency with the other rates, by $k_T^{\mathcal{NP}|\alpha\cdot\delta,\mathcal{U}} = Q/\mu\,k_{\text{basic}}\gamma^{n(\alpha,\delta)}$.

We remark that, due to the definition of $n(\alpha,\delta)$ as the number of neighboring units in permissive state, the units located at the ends of the filament cannot have $n = 2$. Coherently with this, in Eq. (4), the state of the missing neighboring RUs is set to $\mathcal{N}$.

In conclusion, the transition rates are determined by the five parameters $Q$, $\mu$, $k_{\text{d}}$, $k_{\text{off}}$, $k_{\text{basic}}$ (plus the parameter $\gamma$ that regulates the amount of cooperativity), resulting from the eight free parameters constrained by the two assumptions ($\mathcal{U} \to \mathcal{B}$ not affected by Tm, $\mathcal{P} \to \mathcal{N}$ not affected by Tn) and by the detailed-balance consistency.

Concerning the dependence on the filament overlap, we assume that the only transition affected by filament overlap is $\mathcal{N} \to \mathcal{P}$, that is prevented in the central zone of the sarcomere, where the two AFs meet [37]. Specifically, we set, for $\eta \in \{\mathcal{U},\mathcal{B}\}$ and for $\alpha,\delta \in \{\mathcal{N},\mathcal{P}\}$:

$$k_{T,i}^{\mathcal{NP}|\alpha\cdot\delta,\eta} = \chi_{SF}(SL,i)\ k_T^{\mathcal{NP}|\alpha\cdot\delta,\eta}. \tag{25}$$

The resulting 4-states CTMC associated with each RU is represented in Fig. 10.

### 3.3.2 XBs transition rates

On the basis of the results of [82], we work under the hypothesis that the total attachment-detachment rate is independent of the XB elongation. In this case, the models SE-ODE and MF-ODE can be used in place of the more computationally expansive counterparts SE-PDE and MF-PDE, which involve the solution of a PDE system. As a matter of fact, in [82] we have shown that the introduction of such hypothesis significantly reduces the number of parameters to be fitted by experiments, still preserving the capability of the models of reproducing a wide range of experimental characterizations. Moreover, we also make the reasonable assumption that the sliding velocity only affects the detachment rate.

---

* The physical interpretation of the constant $\gamma$ corresponds to $\gamma = \exp(2\frac{\Delta E}{k_B T})$, where $k_B$ is the Boltzmann constant, $T$ the absolute temperature and $\Delta E$ denotes the energetic gain of the configuration of neighboring units in the same state (i.e. $\mathcal{N}$-$\mathcal{N}$ or $\mathcal{P}$-$\mathcal{P}$) with respect to that with different states (i.e. $\mathcal{N}$-$\mathcal{P}$ or $\mathcal{P}$-$\mathcal{N}$). See [85] for more details in this regard.

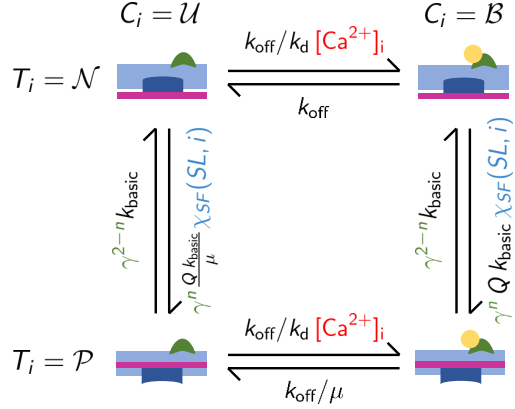

Figure 10: The proposed four states Markov model describing each RU. The terms depending on the intracellular calcium concentration $[\text{Ca}^{2+}]_i$ are highlighted in red; terms depending on the state of neighbouring RUs (i.e. depending on $n$) are highlighted in green; terms depending on the position of the RU and the current sarcomere elongation are highlighted in blue.

As already mentioned, the transition rates may be affected by the mutual arrangement of the filaments. Specifically, we assume that binding is possible only in the single-overlap region and when Tm is in the state $\mathcal{P}$. In other terms, we set:

$$f_{\mathcal{N}}^i(x,v) = 0, \qquad f_{\mathcal{P}}^i(x,v) = f_{\mathcal{P}}(x,v)\,\chi_M(SL,i)\,\chi_{SF}(SL,i). \tag{26}$$

We assume that the unbinding rate can take different values inside and outside the single-overlap region. Hence, we set, for $\alpha \in \{\mathcal{N},\mathcal{P}\}$ and $i \in \mathcal{I}_A$:

$$g_\alpha^i(x,v) = g_\alpha(x,v)\,\chi_M(SL,i)\,\chi_{SF}(SL,i) + \tilde{g}_\alpha(x,v)\,(1 - \chi_M(SL,i)\,\chi_{SF}(SL,i)). \tag{27}$$

Moreover, it is well motivated that out of the single-overlap zone, the detachment rates are not affected by the state of Tm (i.e. $\tilde{g}_{\mathcal{N}} \equiv \tilde{g}_{\mathcal{P}}$) and that the detachment rate when Tm is in state $\mathcal{N}$ is not affected by the filament overlap (i.e. $\tilde{g}_{\mathcal{N}} \equiv g_{\mathcal{N}}$). In summary, we have:

$$f_{\mathcal{P}}(x,v) + g_{\mathcal{P}}(x,v) = g_{\mathcal{N}}(x,v) = \tilde{g}_{\mathcal{N}}(x,v) = \tilde{g}_{\mathcal{P}}(x,v) = r_0 + q(v),$$

for some constant $r_0$ and function $q$, such that $q(0) = 0$. Moreover, since we focus on the small velocity regimes, the function $q$ is well characterized by its behavior around $v = 0$. Hence, we set $q(v) = \alpha|v|$, in order to ease the calibration process.

We notice that, from (10) and (26), it follows, for $p = 0, 1$:

$$\mu_{f_{\mathcal{P}}^i}^p = \mu_{f_{\mathcal{P}}}^p\,\chi_M(SL,i)\,\chi_{SF}(SL,i). \tag{28}$$

Therefore, the parameters to calibrate are the scalars $\mu_{f_{\mathcal{P}}}^0$, $\mu_{f_{\mathcal{P}}}^1$, $r_0$, $\alpha$ and $a_{\text{XB}}$, to link the microscopic force with the macroscopic active tension.

## 4 Parameters calibration

We developed a pipeline to calibrate the parameters of the models proposed in this paper from measurements typically available from experiments. Our strategy is based on the observation that different experimental setups involve different time scales and different aspects of the force generation phenomenon. This allows to decouple the role of parameters associated with different phenomena and with different time scales, thus calibrating them in sequential manner. The calibration pipeline is made of three building blocks: the calibration of the XB rates, the calibration of the RU rates ruling the steady-state solution and, finally, the calibration of the RU rates ruling the kinetics of force development and relaxation. We report below the main steps of the calibration pipeline. A detailed description of the different steps is available in App. B.

## 4.1 Calibration of the XBs rates

Even if the thin-filament activation precedes the XB cycling from a logical viewpoint, we start by illustrating the calibration procedure for the latter part. The reason will be clarified later. In [82] we have shown that the parameters of the distribution-moments equations describing the XB dynamics can be calibrated to fit the steady state force, the shape of the force-velocity relationship (see Fig. 2b) and the slope of the tension-elongation curve following a fast step (see Fig. 2c). We remark that the experimental setups associated with these measurements are are such that the thin filament activation machinery can be considered in steady-state. This observation is crucial since it allows to decouple the calibration of the parameters involved in the thin filament regulation from the calibration of the parameters involved in XBs cycling.

In conclusion, once the parameters of the thin filament activation model has been calibrated, we have at our disposal an automatic procedure to calibrate the remaining parameters. For this reason, we first setup such calibration procedure for the parameters associated with XB cycling (i.e. (11) or (21)) and, successively, we calibrate the parameters associated with RU activation (i.e. (5) or (15)), so that we can directly see the effect of changes of such parameters on the resulting force (the remaining parameters are automatically adjusted).

## 4.2 Calibration of the RUs rates (steady-state)

The steady-state solution of the thin filament activation models (i.e. (5) and (15)) only depends on the ratio between the pairs of opposite transition rates (e.g. the ratio $k_T^{\mathcal{NP}|\alpha\cdot\delta,\mathcal{B}}/k_T^{\mathcal{PN}|\alpha\cdot\delta,\mathcal{B}} = Q\,\gamma^{2n(\alpha,\delta)-2}$). Therefore, the six parameters can be split into two groups: the first group ($Q$, $\mu$, $k_\mathrm{d}$ and $\gamma$) determines the steady-state solution, while the second group ($k_\mathrm{off}$, $k_\mathrm{basic}$) only affects the kinetics of the model (that is to say how fast the transients are). This allows to calibrate first the parameters of the first group, and, only successively, those of the second group.

The fingerprint of the steady-state solution of the RU model is the force-calcium relationship (see Fig. 2a). Hence, we tune the parameters $Q$, $\mu$, $k_\mathrm{d}$ and $\gamma$ to fit this curve. Specifically, $k_\mathrm{d}$ mainly acts on the calcium sensitivity (i.e. $EC_{50}$), $\gamma$ on the apparent cooperativity (i.e. $n_H$), while $Q$ and $\mu$ affect cooperativity, calcium sensitivity, the asymmetry of the force-calcium relationship below and above $EC_{50}$ [90] and the $SL$-driven regulation on calcium sensitivity (in the SE-ODE model).

As we show in Sec. 5, the force-calcium curves obtained with the SE-ODE model exhibit the $SL$-induced change in calcium sensitivity observed in experiments (LDA, see Sec. 2.1). We remark that we are here able to reproduce the LDA without any phenomenological $SL$-dependent tuning of the parameter, as done, e.g. in [67, 107–109]. The LDA emerges from the SE-ODE model in a spontaneous way, as a consequence of the spatial-dependent interaction between the RUs (see [85] for a discussion on this topic). Conversely, with the MF-ODE model it is not possible to reproduce the LDA by simply acting on the model parameters. Indeed, the only effect of $SL$ in the model is to multiplicatively tune the generated force by the factor $\chi_{\mathrm{so}}(SL(t))$. Therefore, no $SL$ induced effect on the calcium sensitivity can be achieved. The mechanism reproducing LDA in the SE-ODE model is indeed intrinsically linked to the explicit spatial representation of the myofilaments [85]. Therefore, in the mean-field model MF-ODE, without an explicit spatial representation, we phenomenologically tune the calcium sensitivity $k_{\mathrm{d}}$ in function of $SL$, by setting

$$k_{\mathrm{d}}(t) = \overline{k}_{\mathrm{d}} + \alpha_{k_{\mathrm{d}}}(SL(t) - SL_{k_{\mathrm{d}}}), \qquad (29)$$

where $SL_{k_{\mathrm{d}}} = 2.15\,\mathrm{\mu m}$.

## 4.3 Calibration of the RUs rates (kinetics)

To complete the calibration of the SE-ODE and MF-ODE models, we only need to set the parameters $k_{\mathrm{basic}}$ and $k_{\mathrm{off}}$, ruling the rapidity at which the transitions $\mathcal{N} \rightleftharpoons \mathcal{P}$ and $\mathcal{U} \rightleftharpoons \mathcal{B}$ take place, respectively. Despite the fact that, at this stage, we need to calibrate just two parameters, this reveals some difficulties, mainly related to the following two aspects. First, the interplay between the two parameters is tight and their roles cannot be easily decoupled [67, 74, 75]. This results in a poor identifiability of the parameters: different combinations of parameters give similar results in terms of force transients. This issue has been reported also by [105], while calibrating the models of [62] and [67]. Additionally, the force transients predicted by the model are very sensitive to the calcium transient at input (this is a typical feature of activation models, see e.g. [105]). Therefore, since the experimentally measured calcium transients are much affected by noise (see e.g. [4, 49, 105], a calibration based on the best fit of the model response to experimentally measured calcium transients should be performed with care.

Based on the former remarks, we calibrate the parameters $k_{\mathrm{basic}}$ and $k_{\mathrm{off}}$ by the following procedure. We consider force transients experimentally measured during isometric twitches and synthetic calcium transients fitted from experimentally measured ones. Then, we perform a MC sampling of the parameters $k_{\mathrm{basic}}$ and $k_{\mathrm{off}}$ within prescribed ranges, and we select those values for which the force transients predicted by the model best fit the experimental ones.

## 4.4 Calibration from rat and human experimental data

Due to the lack of a sufficiently large set of measurements from human cells at body temperature [61] to adequately fit all the model parameters, to calibrate the SE-ODE and MF-ODE models for body-temperature human cardiomyocytes we proceed as follows. First, we calibrate the model parameters from rat experiments at room temperature (for which available data are more abundant) and then we adjust the parameters that are reasonably affected by the two varying factors (i.e. inter-species variability and temperature) to fit the available data from human cells as body temperature. We compensate in this way for the data deficiency. We notice that we work under the hypothesis that inter-species variability does not affect the fundamental processes of tissue activation and force generation, but, since different species express different isoforms of the same protein, it can be encompassed by changing the parameters of the same mathematical model (see [105] for a detailed discussion on this topic).

|  | | SE-ODE | | MF-ODE | |
| --- | --- | --- | --- | --- | --- |
| Parameter | Units | Rat, room temp. | Human, body temp. | Rat, room temp. | Human, body temp. |
| **RU steady-state** | | | | | |
| $\mu$ | - | 10 | 10 | 10 | 10 |
| $\gamma$ | - | 20 | 20 | 12 | 12 |
| $Q$ | - | 3 | 3 | 2 | 2 |
| $\bar{k}_\mathrm{d}$ | µM | 1.622 | 0.74 | 0.835 | 0.381 |
| $\alpha_{k_\mathrm{d}}$ | µM µm$^{-1}$ | 0 | 0 | $-1.258$ | $-0.571$ |
| **RU kinetics** | | | | | |
| $k_\mathrm{off}$ | s$^{-1}$ | 120 | 100 | 120 | 100 |
| $k_\mathrm{basic}$ | s$^{-1}$ | 24 | 13 | 20 | 13 |
| **XB cycling** | | | | | |
| $\mu_{f\mathcal{P}}^0$ | s$^{-1}$ | 57.416 | 57.157 | 32.708 | 32.653 |
| $\mu_{f\mathcal{P}}^1$ | s$^{-1}$ | 1.368 | 1.362 | 0.779 | 0.778 |
| $r_0$ | s$^{-1}$ | 134.31 | 134.31 | 134.31 | 134.31 |
| $\alpha$ | - | 25.184 | 25.184 | 25.184 | 25.184 |
| **Micro-macro upscaling** | | | | | |
| $a_\mathrm{XB}$ | MPa | 22.894 | 22.894 | 22.894 | 22.894 |

Table 3: Parameters of the SE-ODE and MF-ODE models calibrated for room-temperature rat and body-temperature human cells.

| Parameter | Value | Units | Parameter | Value | Units |
| --- | --- | --- | --- | --- | --- |
| $SL_0$ | 2.2 | µm | $\varepsilon$ | 0.05 | µm |
| $L_A$ | 1.25 | µm | $N_M$ | 18 | - |
| $L_M$ | 1.65 | µm | $N_A$ | 32 | - |
| $L_H$ | 0.18 | µm | | | |

Table 4: List of geometrical constants.

Specifically, different species mainly differ in their calcium-sensitivity (i.e. $k_\mathrm{d}$) and in the kinetics (different species feature highly different heart rates), while temperature mainly affects the kinetics [7, 39, 50]. By exploiting the decoupling of the parameters of the RUs model ruling the steady-state relationships from those ruling the kinetics (see Sec. 4.3), we first focus on the steady-state, and we adjust $k_\mathrm{d}$ to reflect the higher calcium sensitivity of human cells compared to rat [61, 67, 105]. Then, we recalibrate the parameters $k_\mathrm{off}$ and $k_\mathrm{basic}$ based on the kinetics of human force transients experimentally measured from human cells at body temperature.

We provide in Tab. 3 the full list of parameters for both species (room-temperature rat and body-temperature human) and for both models (SE-ODE and MF-ODE). Finally, we report in Table 4 the geometrical constants describing the size of the myofilament components that we use in the following.

# 5 Numerical results

We show the results of numerical simulations obtained with the SE-ODE and MF-ODE models, performed under different experimental settings. The numerical schemes

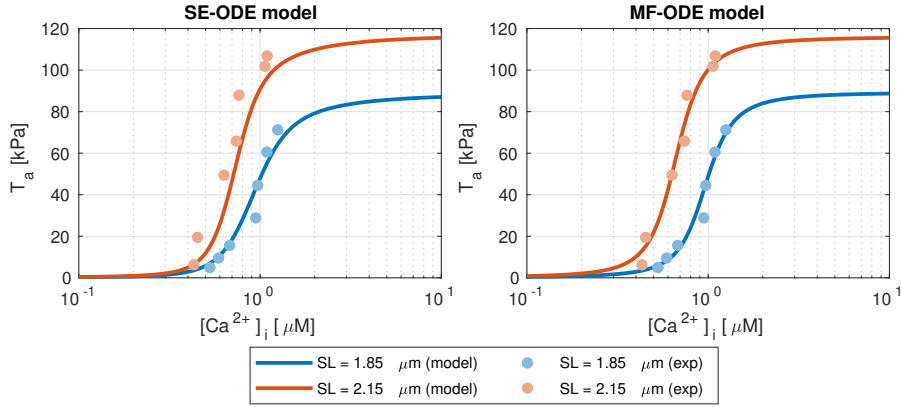

Figure 11: Steady-state force-calcium curves obtained with the SE-ODE (left) and MF-ODE (right) models with the optimal parameters values reported in Tab. 3 for $SL = 1.85\,\mu\text{m}$ and $SL = 2.15\,\mu\text{m}$, compared with experimental data from intact rat cardiac cells at room temperature, from [101].

employed to approximate the solutions of the models here proposed are described in App. C. The codes implementing the models proposed in this paper and used to produce the results presented in this section are freely available in the following online repository:

https://github.com/FrancescoRegazzoni/cardiac-activation

With this implementation, the computational cost of the SE-ODE model is of nearly 12 s of simulation for 1 s of physical time on a single-core standard laptop. The MF-ODE model, instead, requires nearly 1 s of computational time to simulate 1 s of physical time on the same computer platform. Moreover, all the data used to generate the figures of this paper are available in the static repository [84]:

https://doi.org/10.5281/zenodo.3992553

## 5.1 Steady-state results

First, we consider steady-state solutions. To numerically obtain the steady-state curves, we fix a level of $[\text{Ca}^{2+}]_\text{i}$ and $SL$ and we let the model reach the equilibrium solution.

### 5.1.1 Force-calcium relationship

We report in Fig. 11 the force-calcium curves obtained with the SE-ODE and MF-ODE models calibrated from room-temperature rat data, compared with the experimental data used for the calibration. We are able to well fit the main features of the curves, including the characteristic S-shape, the plateau forces at both $SL$, the significant cooperativity typical of the cardiac tissue and the $SL$-induced change in calcium sensitivity. In Fig. 12 we report the steady-state curves obtained with the sets of parameters calibrated for human body temperature cells. Also in this case, the curves reproduce the main experimentally observed features reported in Sec. 2.1.

Figure 13 shows the dependence of the Hill coefficient $n_H$ and of the half-activating calcium concentration $\text{EC}_{50}$ on the sarcomere length $SL$. We notice that, while the MF-ODE model produces an Hill coefficient that is independent of $SL$ (the reason is that the role of $SL$ on activation is just that of shifting and rescaling the curves, thus leaving $n_H$ unaffected), the SE-ODE model predicts a small increase of $n_H$ with $SL$. Both the results are equally acceptable since there is no common agreement on whether the Hill coefficient depend on $SL$ or not [27, 51, 55, 101, 102]. Both the models correctly predict for both species an increase of $EC_{50}$ as $SL$ decreases. The relationship is approximately linear in the typical working range of $SL$ (as experimentally observed, e.g., in [27]), while, for small values of $SL$, the SE-ODE model produces a faster decrease of sensitivity.

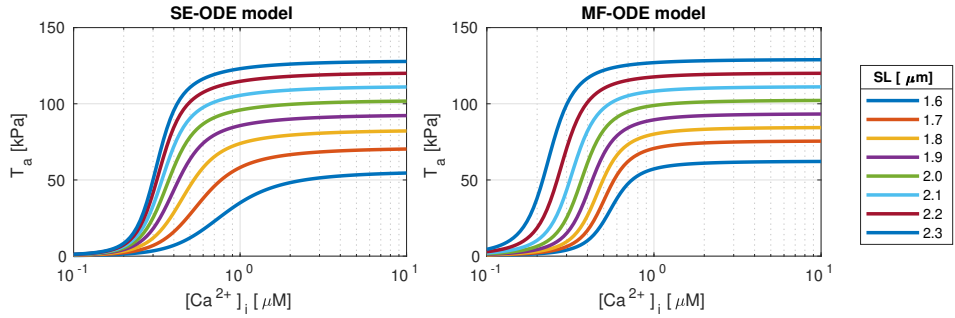

Figure 12: Steady-state force-calcium relationship at different $SL$ obtained with the SE-ODE (left) and the MF-ODE (right) models for intact, body-temperature human cardiomyocytes.

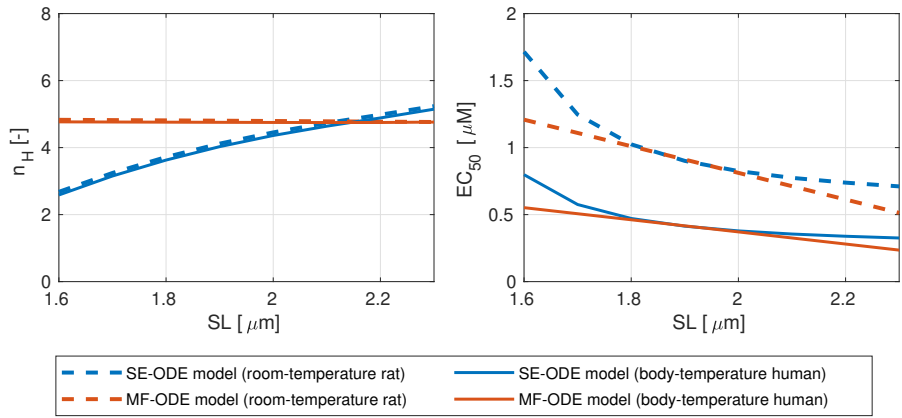

Figure 13: Dependence of the Hill coefficient $n_H$ and of the half-activating calcium concentration $EC_{50}$ on the sarcomere length $SL$, for the SE-ODE (blue lines) and MF-ODE (red lines) model, calibrated for intact, room-temperature rat cardiomyocytes (dashed lines) and for intact, body-temperature human cardiomyocytes (solid lines).

### 5.1.2 Force-length relationship

Figure 14 shows the ascending limb of the steady-state force-length relationship. For both the SE-ODE and MF-ODE models we observe a change of slope for saturating calcium concentration around 1.65 μm, coherently with the experimental observations [54, 56, 101, 102]. Moreover, both models predict the observed change of convexity of the force-length curves at different calcium levels [55, 101, 102].

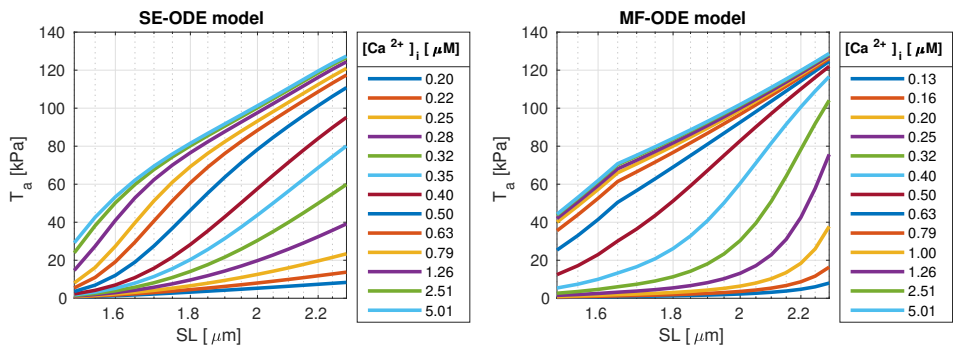

Figure 14: Steady-state force-length relationship at different $[Ca^{2+}]_i$ obtained with the SE-ODE (left) and the MF-ODE (right) models for intact, body-temperature human cardiomyocytes.

## 5.2 Isometric twitches

The predicted isometric twitches obtained with the room-temperature rat models are compared with the experimental data used for their calibration in Fig. 15. The calcium transient here employed is obtained by fitting the experimental transient of [49] with the synthetic curve reported in App. B.

Similarly, we show in Fig. 16 the tension transients obtained by simulating isometric twitches giving as input to the human models the calcium transient of the ToR-ORd model [104]. We notice that both models predict the tension-dependent prolongation of the twitch time [4, 48, 49], as it can be seen from the normalized traces reported in the bottom lines of the figures. We remark that, despite recent measurements on rabbit cells show that part of the increase of twitch force is linked to a larger calcium release under stretch [10], in this paper – due to the lack of experimental data showing a similar effect in human cells – we employ for simplicity the same calcium transient for all lengths.

In order to quantitatively assess the effects of $SL$ on twitches, we report in Fig. 17 the dependence on $SL$ of the peak force ($T_a^{\text{peak}}$) and of some synthetic indicators of the kinetics. Specifically, we consider the time-to-peak $TTP$ (defined as the time separating the beginning of the stimulus and the tension peak) and the relaxation times $RT_{50}$ and $RT_{90}$ (defined as the time needed to accomplish 50% and 90% of relaxation, respectively). We notice that both models feature a kinetics with characteristic times matching those obtained experimentally. Moreover, both the peak tension and the characteristic times feature the expected increasing behavior with respect to $SL$. Remarkably, in the SE-ODE model, this feature spontaneously emerges from the model. Conversely, in the MF-ODE model, the prolongation of $TTP$ and of the relaxation times is to be ascribed to the change of calcium sensitivity with respect to $SL$. Furthermore, it is remarkable that, although in both the cases the kinetic constants $k_{\text{off}}$ and $k_{\text{basic}}$ have been calibrated to fit the twitch kinetics for a given $SL$, the simulations yields a good experimental match also for other values of $SL$ (this happens despite the $SL$-dependent effect on calcium sensitivity is calibrated under a different experimental setting, i.e. steady-state conditions).

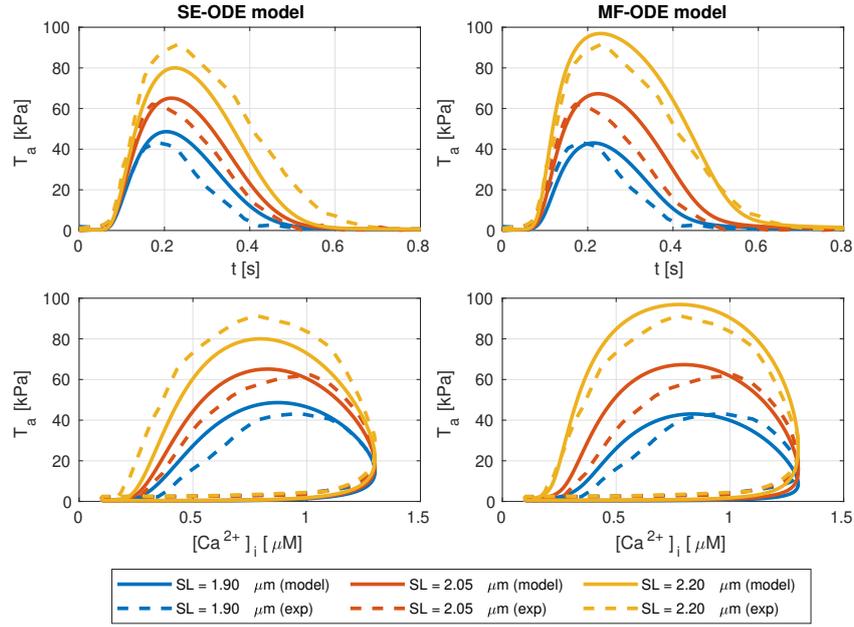

Figure 15: Force transients (top line) and phase-loops (bottom line) in isometric twitches, for different values of $SL$, predicted by the SE-ODE model (left column) and MF-ODE model (right column), in comparison with the experimental measurements from intact rat cardiac cells taken from [48].

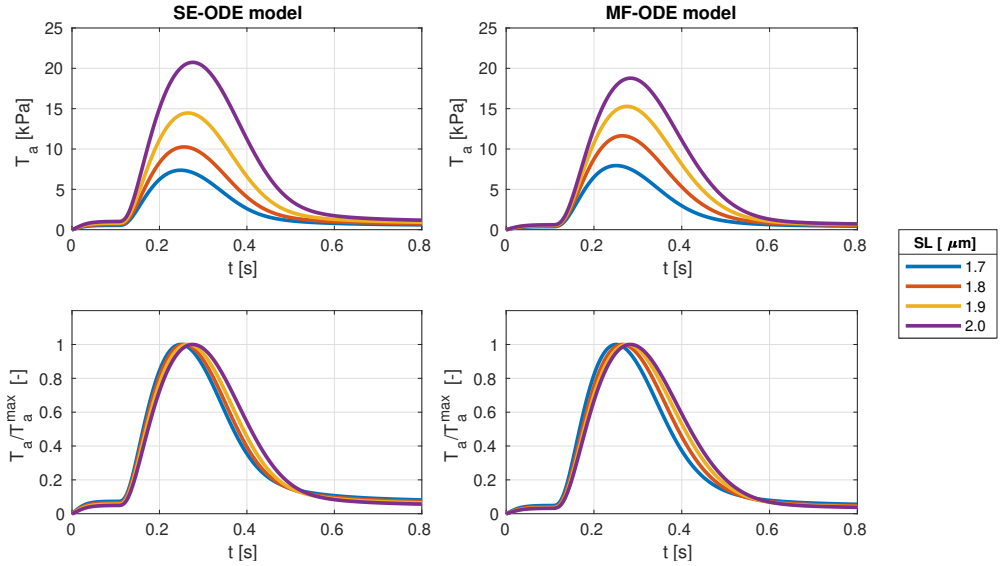

Figure 16: Tension transients during isometric twitches at different $SL$ obtained with the SE-ODE (left) and the MF-ODE (right) models for intact, body-temperature human cardiomyocytes.

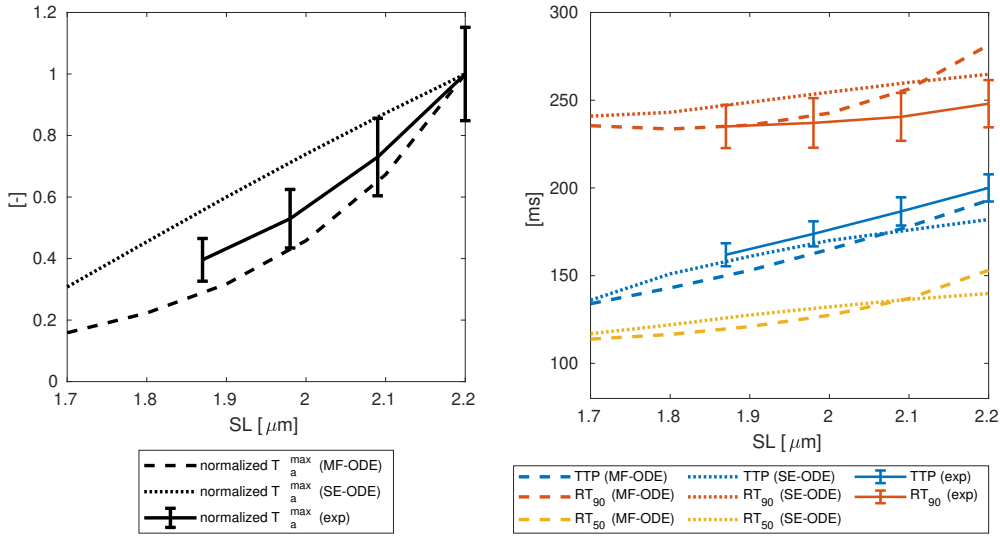

Figure 17: Tension peak, normalized with respect to the value at 2.2 μm (left), and metrics of activation and relaxation kinetics (right) as function of $SL$ during isometric twitches obtained with the SE-ODE and the MF-ODE models for intact, body-temperature human cardiomyocytes. When possible, model results are compared with experimental data from [18].

### 5.3 Force-velocity relationship

Figure 18 shows the force-velocity relationship predicted with the rat models, compared with the experimental data used for the calibration. The human model yields similar results. For both the SE-ODE and the MF-ODE model, the numerical results correctly reproduce the experimentally observed convex profile, with the force reaching zero in correspondence of a finite value of velocity, the so-called maximum shortening velocity (see Sec. 2.2). Moreover, the value of $v_{hs}^{max}$ is not significantly affected by the level of activation (that is to say, by the values of $[Ca^{2+}]_i$ and $SL$), as well as the curvature of the curve. This is also coherent with the experimental observations [12]. The good agreement with the experimental data serves as a validation for the automatic calibration procedure presented in Sec. 4.1 .

### 5.4 Fast force transients

We consider the response to fast steps predicted by the SE-ODE and the MF-ODE models. With this aim, we set a fixed value for the calcium concentration and sarcomere length (we set $[Ca^{2+}]_i = 1.2$ μM and $SL = 2.2$ μm, but the results are not significantly affected by this choice) and we let the system reach the steady-state. Then, we apply a length step, by applying a constant shortening velocity in a small time interval $\Delta t = 200$ μs, and we plot the tension at the end of the step as a function of the step length $\Delta L$.

We show in Fig. 19 the results obtained with the rat models, with a comparison with experimental data (similar results are obtained with the human models). The good match between the simulation results and the experimental measurements provide a further validation of the calibration procedure.

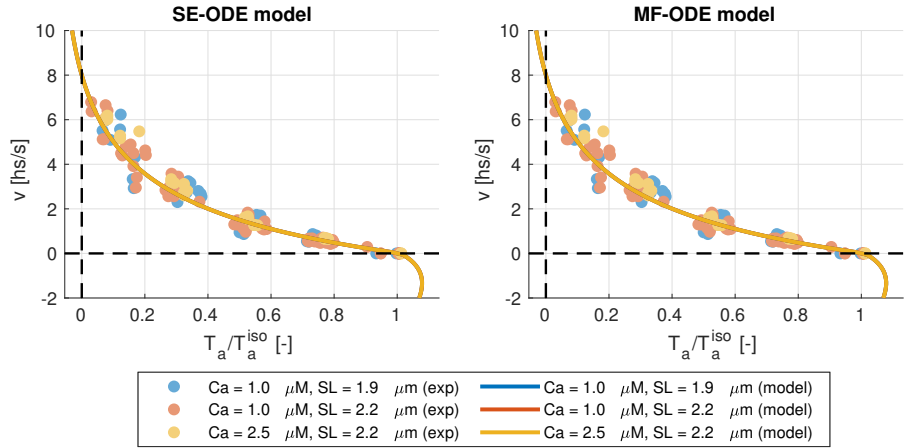

Figure 18: Normalized force-velocity relationships for different combinations of $[\text{Ca}^{2+}]_i$ and $SL$ obtained with the SE-ODE (left) and the MF-ODE (right) models for intact, room-temperature rat cardiomyocytes in comparison with experimental measurements from [12].

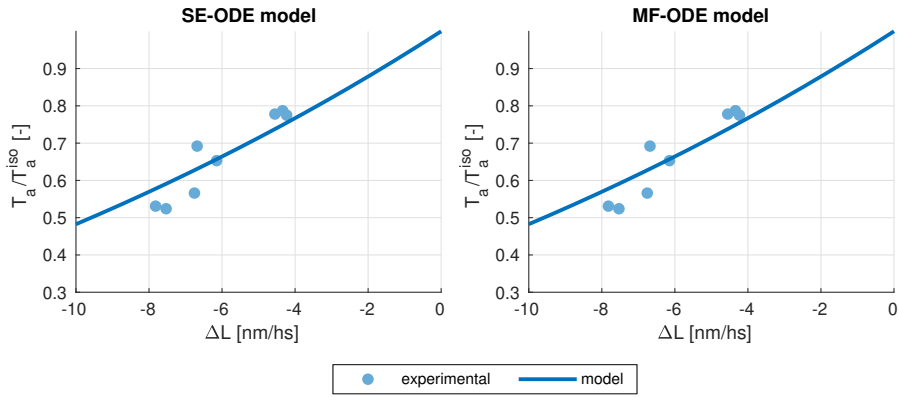

Figure 19: Normalized force after the application of a fast length step $\Delta L$ for intact, room-temperature rat cardiomyocytes. Solid line: model result; circles: $T_2$-$L_2$ experimental data from [12].
32

## 5.5 Multiscale cardiac electromechanics

Finally, we consider a multiscale cardiac EM model of the left ventricle (LV), that we describe in the following. Further details on the EM model and its numerical discretization can be found in [85]. For the sake of brevity, we only show the results obtained by considering the MF-ODE model for human body-temperature cardiomyocytes.

We employ a computational domain $\Omega_0$ derived from the Zygote heart model, representing the $50^{\text{th}}$ percentile of a 21 years old healthy caucasian male [113], in which we generate the fibers and sheets distribution by the rule-based algorithm proposed in [6]. We model the electrophysiological activity of the heart tissue by means of the monodomain equation [21, 22], coupled with the TTP06 model for the ionic activity [99]. We introduce a deformation map $\varphi\colon \Omega_0 \times [0,T] \to \mathbb{R}^3$ and we define the displacement vector as $\mathbf{d}(\mathbf{X},t) = \varphi(\mathbf{X},t) - \mathbf{X}$. The mechanical behavior of the myocardium is described by means of the balance of momentum equation for the displacement $\mathbf{d}$ [2, 70], where we model the passive properties of the cardiac tissue through the quasi-incompressible exponential constitutive law of [106]. To account for the presence of the pericardium, we employ the generalized Robin boundary conditions of [35, 73] on the epicarial boundary. On the LV base we adopt the energy-consistent boundary condition that we proposed in [83], accounting for the effect of the neglected part of the domain on the artificial boundary of the LV base.

Within a multiscale setting, we describe the force generation mechanisms at the microscale by means of the MF-ODE model, proposed in this paper. The intracellular calcium concentration ($[\text{Ca}^{2+}]_\text{i}$) is provided in each point of the computational domain by the TTP06 model. The local sarcomere length, in turn, is assumed to be proportional to the tissue stretch in the direction of muscel fibers $\mathbf{f}_0$, that is we set $SL = SL_0 \sqrt{\mathcal{I}_{4,f}}$, where $SL_0$ denotes the sarcomere length at rest, $\mathcal{I}_{4,f} = \mathbf{F}\mathbf{f}_0 \cdot \mathbf{F}\mathbf{f}_0$ and $\mathbf{F} = \mathbf{I} + \nabla \mathbf{d}$ denotes the deformation gradient. The last input of the MF-ODE model, the normalized shortening velocity, is obtained by definition as $v = -\frac{\partial}{\partial t} SL / SL_0$.

The output of the MF-ODE model, namely the active tension magnitude field $T_\text{a}$, provides the link from the microscopic to the macroscopic level, where the effect of the microscopically generated active force is given by the following active stress tensor [36, 85]:
$$\mathbf{P}^{\text{act}} = T_\text{a} \frac{\mathbf{F}\mathbf{f}_0 \otimes \mathbf{f}_0}{\|\mathbf{F}\mathbf{f}_0\|}.$$

Finally, the 3D LV model is coupled to a 0D model of blood circulation, consisting of a two-elements Windkessel model [110].

For the spatial discretization of the equations involved in the EM model, we employ piece-wise linear Finite Elements of a tetrahedral computational mesh with $354 \cdot 10^3$ cells. For the discretization of time derivatives, we use first-order finite difference schemes [78] with the implicit-explicit (IMEX) scheme described in [85]. The coupling of the different core models is obtained by means of the segregated strategy presented in [26].

The results of the numerical simulation are provided in Fig. 20, where we show the deformation of the LV at different time steps of an heartbeat, and in Fig. 21, showing the time evolution of the variables involved in the force generation phenomenon and of the LV pressure and volume predicted by the multiscale EM model. In Fig. 22 we show the pressure-volume loops obtained for different preloads, i.e. by varying the end-diastolic pressure $p_{\text{ED}}$. Our model correctly predicts the increased stroke volume following a raise in the preload. This phenomenon, known as Frank-Starling effect, represents a self-regulatory mechanism of fundamental importance, as it allows the cardiac output to be synchronized with the venous return [52]. The microscopic driver of the Frank-Starling effect is the length-dependent mechanisms of force generation. We remark that, in our EM model, the Frank-Starling is not artificially imposed, but it spontaneously emerges from the properties of the microscopic model of force generation.

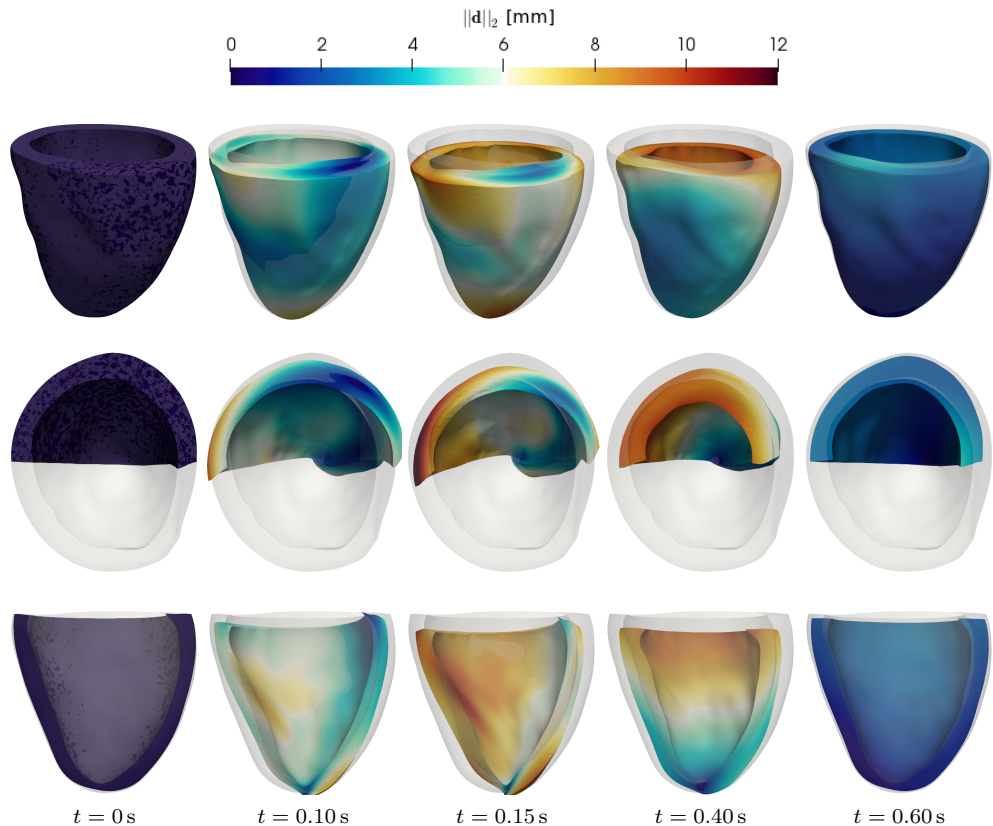

Figure 20: LV multiscale EM: deformed geometry and magnitude of displacement **d** at different times. Top row: full geometry. Middle row: half domain (top view). Bottom row: half domain (frontal view).

# 6  Conclusions

In this paper we have derived four different models for the generation of active force in the cardiac muscle tissue. Such models are based on a biophysically detailed description of the microscopic mechanisms of regulation and activation of the contractile proteins. Still, their numerical realization features a contained computational cost. Indeed, their numerical resolution does not require the computationally expensive MC method.

The main difficulties to be addressed in the derivation of full-sarcomere models concern (i) the spatial correlation of the states of the RUs due to the nearest-neighbor interactions of Tm, which hinders a straightforward decoupling of the adjacent RUs, and (ii) the mutual filament sliding, that changes which RU regulates which XB from time to time. Similarly to what we did in [81], we address the first problem by introducing a conditional independence assumption for far RUs, given the state of intermediate RUs (this is coherent with the local nature of nearest-neighboring interactions). This assumption dramatically reduces the number of equations needed to describe the evolution of the stochastic processes: while in the original model of [90] the number of variables of the CTMC increases exponentially with the number of RUs, in our model the number of variables grows linearly with the number of RUs.

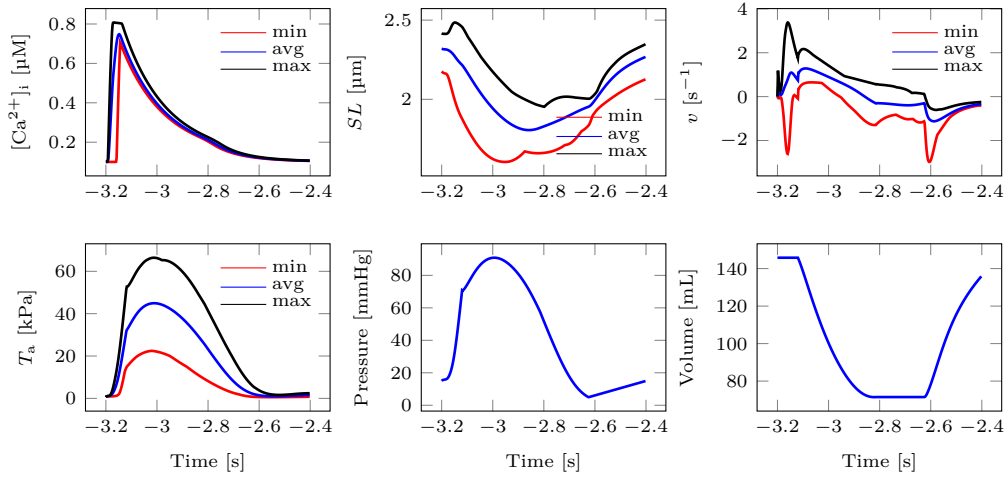

Figure 21: LV multiscale EM. Time evolution of $[Ca^{2+}]_i$, $SL$, $v$ and $T_a$ (minimum, maximum and average over the computational domain) and of left ventricle pressure and volume.

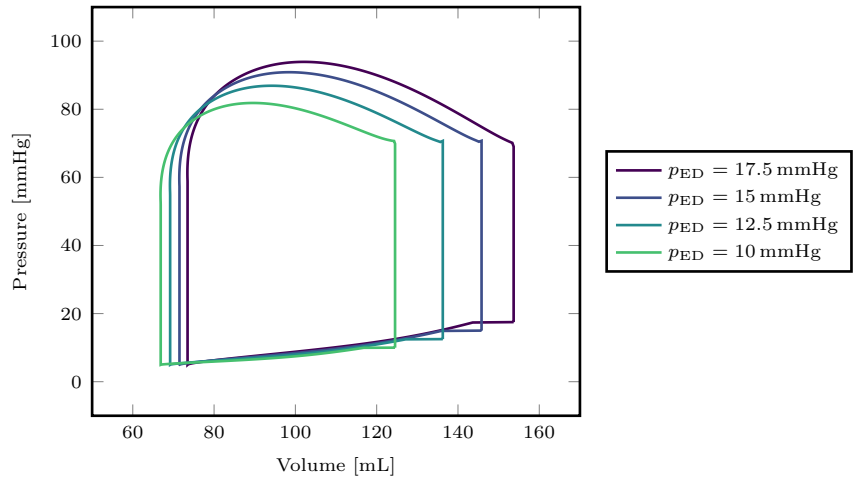

Figure 22: Pressure-volume loops obtained with different preloads (reported in legend).

Since no feedback from the XBs to the RUs is foreseen, the dynamics of the latter can be considered independently of that of the former. Moreover, we depart from the traditional MHs-centered representation of XBs, in favour of a BSs-centered point of view. Thanks to this change of perspective, we derived a set of equations describing the XB dynamics without the need to track the mutual position of the RUs and the MHs.

Under the hypothesis that the total attachment-detachment rate is independent of the myosin arm stretch (as done in [8, 17]), the PDE system describing the XBs can be replaced by a system of ODEs. We remark that this is not a simplificatory assumption, like the conditional independence assumptions mentioned before, but rather a feature of the specific modeling choice for the transition rates describing the attachment-detachment process.

We have also presented a class of models (MF-PDE and MF-ODE), such that the myofilaments overlap is not explicitly described, but is rather replaced by a mean-field description of a single representative RUs triplet. We remark that such mean-field models differ from the mean-field models of [63, 80, 88, 91, 92]. The latter, indeed, consider a single RU, instead of a triplet. In this manner, the short-range spatial correlation, responsible of cooperativity, cannot be captured. Conversely, the mean-field triplet framework here proposed, thanks to the local nature of cooperativity, allows to capture the effect of nearest-neighbor interactions, as testified by the remarkably good agreement between model predictions as experimental measurements, in particular in the reproduction of the highly cooperative steady-state force-calcium curves (see Sec. 5.1). We have then calibrated the SE-ODE and the MF-ODE models for room-temperature rat intact cardiomyocytes and, later, body-temperature human intact cardiomyocytes.

The SE-ODE model predicts the so-called LDA (the increment of calcium sensitivity when the sarcomere length increases), phenomenon whose microscopic source is still debated [1, 30, 68, 72, 87, 94, 103]. Interestingly, in our SE-ODE model, the LDA spontaneously emerges without any phenomenological tuning the parameters in dependence of $SL$, thanks to the explicit spatial representation of the spatially-dependent nearest-neighborhood interactions among RUs. In particular, we believe that this is linked to the RUs located at the end-points of the single-overlap zone, characterized by a low probability of being in the $\mathcal{P}$ state (because they have at most one neighbor in state $\mathcal{P}$). Since nearest-neighbor interactions propagate the low probability of $\mathcal{P}$ towards the center of the filament, small values of $SL$ are penalized with respect to large values of $SL$, the role of such border effect being more pronounced (further details on this topic can be found in [85]). Therefore, if the hypothesis that the RUs located at the end-point of the single-overlap zone behave as if the outer neighboring units are in state $\mathcal{N}$ is accepted, then the SE-ODE model provides a possible explanation for LDA. Conversely, if this hypothesis in not accepted, then the effect of the edges can be neglected and the mean-field approximation underlying the MF-ODE model is well motivated. In conclusion, according to the hypothesis made on the behavior of the RUs near the end-points of the single-overlap zone, the SE-ODE and the MF-ODE models represent two alternative descriptions of the sarcomere dynamics based on a microscopically detailed representation of the regulatory and contractile proteins, where phenomenological modeling choices are only introduced for phenomena whose underlying mechanisms are not fully understood [1, 68, 94, 103]. The models proposed in this paper could be enriched to investigate alternative hypotheses for the LDA (such as the force-dependent recruitment of MHs from an "off" state, in which the interaction with the BSs is prevented [10]).

The results of the numerical simulations showed that our models can reproduce the main features of the experimental characterizations of muscle contraction associated with the time scales of interest of cardiac EM (that is to say, the time scales larger than that of the power-stroke), including the steady-state relationship between calcium and force and between sarcomere length and force, the kinetics of activation of relaxation, the length-dependent twitches prolongation and peak force increase, and the force-velocity relationship. Additionally, we plan to study the response of the models to changes in the heart rate, which has been reported to have an impact on calcium sensitivity, developed force and twitch kinetics [19, 20, 23]. However, this investigation requires a model for calcium dynamics capable of accounting for the effects of the heart rate on the calcium transients [23].

Finally, we have presented the results of multiscale EM numerical simulations in a human LV, obtained by modeling the microscopic generation of active force through the MF-ODE model, able of reproducing realistic pressure-volume loops. Moreover, our multiscale EM model is capable of reproducing – at the macroscale – the Frank-Starling self-regulation mechanism, in virtue of the length-dependent effects characterizing – at the microscale – the force-generation model. This macroscopic effect, emerging from a microscopic phenomenon, can be regarded as a proof of concept for potential uses of the models proposed in this paper in the context of 3D numerical simulations. The latter, indeed, can help to investigate the links between microscopic mechanisms and organ-level phenomena and to elucidate the relationships intercurrent between the microscopic variables and the macroscopic ones. Furthermore, the employment of a biophysically detailed model in organ-level simulations might reveal the links between cardiomyopathies and their cellular or molecular basis.

## Acknowledgements

This project has received funding from the European Research Council (ERC) under the European Union's Horizon 2020 research and innovation programme (grant agreement No 740132, iHEART - An Integrated Heart Model for the simulation of the cardiac function, P.I. Prof. A. Quarteroni).

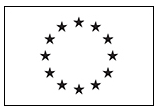
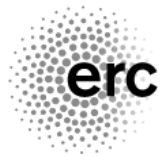

## A  Models derivation

We report the detailed derivation of the models proposed in Sec. 3. The derivation is based on the assumptions there discussed. These assumptions allow to neglect second-order interactions among the stochastic processes, so that the variables can be partially decoupled, thus leading to drastic reductions in the size of model. Such strategy is illustrated in the following proposition.

**Proposition 1.** *Let $(\Omega, \mathcal{A}, \mathbb{P})$ be a probability space. Let $A, B, C \subset \Omega$ and let $\mathcal{D}$ be a finite partition of $\Omega$, such that:*

*(H1) $A \perp\!\!\!\perp B \,|\, C, D \quad \forall D \in \mathcal{D}$;*

*(H2) $B \perp\!\!\!\perp D \,|\, C \quad \forall D \in \mathcal{D}$.*

*Then, we have:*

$$\mathbb{P}[A|B,C] = \frac{\sum_{D \in \mathcal{D}} \mathbb{P}[A|C,D]\mathbb{P}[C,D]}{\mathbb{P}[C]} = \mathbb{P}[A|C].$$

*Proof.* We have:

$$\mathbb{P}[A|B,C] = \frac{\mathbb{P}[A,B,C]}{\mathbb{P}[B,C]} = \frac{\sum_{D \in \mathcal{D}} \mathbb{P}[A,B,C,D]}{\mathbb{P}[B,C]} = \frac{\sum_{D \in \mathcal{D}} \mathbb{P}[A|B,C,D]\mathbb{P}[B,C,D]}{\mathbb{P}[B,C]}$$
$$= \frac{\sum_{D \in \mathcal{D}} \mathbb{P}[A|B,C,D]\mathbb{P}[B|C,D]\mathbb{P}[C,D]}{\mathbb{P}[B,C]}.$$

From (H1), it follows:
$$\mathbb{P}[A|B,C,D] = \mathbb{P}[A|C,D].$$

Moreover, in virtue of (H2), we have:
$$\mathbb{P}[B|C,D] = \mathbb{P}[B|C] = \mathbb{P}[B,C]/\mathbb{P}[C].$$

By substituting into the above equation, the thesis follows. □

In the following, we will use several times the result of Prop. 1, where (H1) is a modeling choice on the dynamics of the system and (H2) is a simplifying assumption. Specifically, $A$ is the target event, whose probability is the aim of the computation. In many situations, we know the joint probability of $C$ and $B$, whereas the probability of $A$ can be obtained by the joint probability of $C$ and a different event $D$. Proposition 1 allows to pass from $B$ to $D$, by assuming that the knowledge of $B$ does not provide any further information when $C$ and $D$ are known.

## A.1 Definition of transition rates

As a starting point for a rigorous derivation of the proposed models, we provide a precise definition of the transition rates that govern the dynamics of the stochastic processes $T_i^t$, $C_i^t$, $A_i^t$, $M_j^t$ and $Z_i^t$. Specifically, we have, for $\delta \in \{\mathcal{B},\mathcal{U}\}$ and $\alpha, \beta, \eta \in \{\mathcal{P},\mathcal{N}\}$:

$$k_{C,i}^{\delta\overline{\delta}|\beta} = \lim_{\Delta t \to 0} \frac{1}{\Delta t}\mathbb{P}\left[C_i^{t+\Delta t} = \overline{\delta} \,|\, (C_i, T_i)^t = (\delta, \beta)\right],$$

$$k_{T,i}^{\beta\overline{\beta}|\alpha \cdot \eta, \delta} = \lim_{\Delta t \to 0} \frac{1}{\Delta t}\mathbb{P}\left[T_i^{t+\Delta t} = \overline{\beta} \,|\, (T_{i-1}, T_i, T_{i+1}, C_i)^t = (\alpha, \beta, \eta, \delta)\right],$$

$$f_\alpha^i(x, v(t)) = \lim_{\Delta t \to 0} \frac{1}{\Delta t}\mathbb{P}\left[Z_i^{t+\Delta t} = x \,|\, Z_i^t = \emptyset, T_i^t = \alpha,\right.$$
$$\left. \exists j \in \mathcal{I}_M : d_{ij}^t = x + v_{\text{hs}}\Delta t, M_j^t = 0\right],$$

$$g_\alpha^i(x, v(t)) = \lim_{\Delta t \to 0} \frac{1}{\Delta t}\mathbb{P}\left[Z_i^{t+\Delta t} = \emptyset \,|\, Z_i^t = x, T_i^t = \alpha\right],$$

where $v(t)$ (i.e. the normalized shortening velocity) is assumed to be given. In the definition of $f_\alpha^i$, the events conditioning the probability ensure that, at time $t$, the $i$-th BS is not attached and that there exists a non-attached MH at distance $x + v_{hs}\Delta t$ (so that at time $t + \Delta t$ the distance is reduced to $x$).

## A.2 Derivation of Eq. (5)

Le us consider the time increment $\Delta t$ and let us compute the probability $\pi_i^{\alpha\beta\delta,\vartheta\eta\lambda}(t + \Delta t)$. In virtue of the Bayes formula [69], we have:

$$\pi_i^{\alpha\beta\delta,\vartheta\eta\lambda}(t + \Delta t) \overset{\Delta t \to 0}{\sim}$$
$$\mathbb{P}\left[T_{i-1}^{t+\Delta t} = \alpha | (T_{i-1}, T_i, T_{i+1})^t = (\overline{\alpha}, \beta, \delta), (C_{i-1}, C_i, C_{i+1})^t = (\vartheta, \eta, \lambda)\right]\pi_i^{\overline{\alpha}\beta\delta,\vartheta\eta\lambda}(t)$$
$$+\mathbb{P}\left[T_i^{t+\Delta t} = \beta | (T_{i-1}, T_i, T_{i+1})^t = (\alpha, \overline{\beta}, \delta), (C_{i-1}, C_i, C_{i+1})^t = (\vartheta, \eta, \lambda)\right]\pi_i^{\alpha\overline{\beta}\delta,\vartheta\eta\lambda}(t)$$
$$+\mathbb{P}\left[T_{i+1}^{t+\Delta t} = \delta | (T_{i-1}, T_i, T_{i+1})^t = (\alpha, \beta, \overline{\delta}), (C_{i-1}, C_i, C_{i+1})^t = (\vartheta, \eta, \lambda)\right]\pi_i^{\alpha\beta\overline{\delta},\vartheta\eta\lambda}(t)$$
$$+\mathbb{P}\left[C_{i-1}^{t+\Delta t} = \vartheta | (T_{i-1}, T_i, T_{i+1})^t = (\alpha, \beta, \delta), (C_{i-1}, C_i, C_{i+1})^t = (\overline{\vartheta}, \eta, \lambda)\right]\pi_i^{\alpha\beta\delta,\overline{\vartheta}\eta\lambda}(t)$$
$$+\mathbb{P}\left[C_i^{t+\Delta t} = \eta | (T_{i-1}, T_i, T_{i+1})^t = (\alpha, \beta, \delta), (C_{i-1}, C_i, C_{i+1})^t = (\vartheta, \overline{\eta}, \lambda)\right]\pi_i^{\alpha\beta\delta,\vartheta\overline{\eta}\lambda}(t)$$
$$+\mathbb{P}\left[C_{i+1}^{t+\Delta t} = \lambda | (T_{i-1}, T_i, T_{i+1})^t = (\alpha, \beta, \delta), (C_{i-1}, C_i, C_{i+1})^t = (\vartheta, \eta, \overline{\lambda})\right]\pi_i^{\alpha\beta\delta,\vartheta\eta\overline{\lambda}}(t)$$
$$+\mathbb{P}\left[(T_{i-1}, T_i, T_{i+1})^{t+\Delta t} = (\alpha, \beta, \delta), (C_{i-1}, C_i, C_{i+1})^{t+\Delta t} = (\vartheta, \eta, \lambda) | \right.$$
$$\left. (T_{i-1}, T_i, T_{i+1})^t = (\alpha, \beta, \delta), (C_{i-1}, C_i, C_{i+1})^t = (\vartheta, \eta, \lambda)\right]\pi_i^{\alpha\beta\delta,\vartheta\eta\lambda}(t),$$

where, by definition, we have:

$$\mathbb{P}\left[T_i^{t+\Delta t} = \beta | (T_{i-1}, T_i, T_{i+1})^t = (\alpha, \overline{\beta}, \delta), (C_{i-1}, C_i, C_{i+1})^t = (\vartheta, \eta, \lambda)\right] \overset{\Delta t \to 0}{\sim} k_{T,i}^{\overline{\beta}\beta|\alpha \cdot \delta, \eta}\Delta t,$$

and

$$\mathbb{P}\left[C_{i-1}^{t+\Delta t} = \vartheta | (T_{i-1}, T_i, T_{i+1})^t = (\alpha, \beta, \delta), (C_{i-1}, C_i, C_{i+1})^t = (\overline{\vartheta}, \eta, \lambda)\right] \overset{\Delta t \to 0}{\sim} k_{C,i-1}^{\overline{\vartheta}\vartheta|\alpha}\Delta t,$$
$$\mathbb{P}\left[C_i^{t+\Delta t} = \eta | (T_{i-1}, T_i, T_{i+1})^t = (\alpha, \beta, \delta), (C_{i-1}, C_i, C_{i+1})^t = (\vartheta, \overline{\eta}, \lambda)\right] \overset{\Delta t \to 0}{\sim} k_{C,i}^{\overline{\eta}\eta|\beta}\Delta t,$$
$$\mathbb{P}\left[C_{i+1}^{t+\Delta t} = \lambda | (T_{i-1}, T_i, T_{i+1})^t = (\alpha, \beta, \delta), (C_{i-1}, C_i, C_{i+1})^t = (\vartheta, \eta, \overline{\lambda})\right] \overset{\Delta t \to 0}{\sim} k_{C,i+1}^{\overline{\lambda}\lambda|\delta}\Delta t.$$

By adopting assumption (H1) and applying Prop. 1 for $A = (T_{i-1}^{t+\Delta t} = \overline{\alpha})$, $B = (T_{i+1}^t = \delta, C_{i+1}^t = \lambda)$, $C = ((T_{i-1}, T_i)^t = (\alpha, \beta), (C_{i-1}, C_i)^t = (\vartheta, \eta))$ and $\mathcal{D} = \{(T_{i-2}^t = \xi, C_{i-2}^t = \zeta)\}_{\xi,\zeta}$, we have:

$$\mathbb{P}\left[T_{i-1}^{t+\Delta t} = \alpha | (T_{i-1}, T_i, T_{i+1})^t = (\overline{\alpha}, \beta, \delta), (C_{i-1}, C_i, C_{i+1})^t = (\vartheta, \eta, \lambda)\right]$$
$$= \left(\sum_{\xi,\zeta} \pi_{i-1}^{\xi\overline{\alpha}\beta,\zeta\vartheta\eta}(t)\right)^{-1} \sum_{\xi,\zeta} \mathbb{P}\left[T_{i-1}^{t+\Delta t} = \alpha | (T_{i-2}, T_{i-1}, T_i)^t = (\xi, \overline{\alpha}, \beta),\right.$$
$$\left.(C_{i-2}, C_{i-1}, C_i)^t = (\zeta, \vartheta, \eta)\right] \pi_{i-1}^{\xi\overline{\alpha}\beta,\zeta\vartheta\eta}(t)$$
$$= \frac{\sum_{\xi,\zeta} k_{T,i}^{\overline{\alpha}\alpha|\xi \cdot \beta,\vartheta} \pi_{i-1}^{\xi\overline{\alpha}\beta,\zeta\vartheta\eta}(t)}{\sum_{\xi,\zeta} \pi_{i-1}^{\xi\overline{\alpha}\beta,\zeta\vartheta\eta}(t)} \Delta t + o(\Delta t),$$

and similarly for the term related to $T_{i+1}^{t+\Delta t}$. In conclusion, by taking the limit $\Delta t \to 0$, we obtain Eq. (5).

## A.3 Derivation of Eq. (15)

Before showing the derivation of Eq. (15), we precisely state the hypothesis of invariance by translation of the joint distribution of RUs. Specifically, we assume that, for any set of indices $\mathcal{I}_1 \subset \mathbb{Z}$ and $\mathcal{I}_2 \subset \mathbb{Z}$ and for any collection of states $\alpha_i \in \{\mathcal{N}, \mathcal{P}\}$ (for $i \in \mathcal{I}_1$) and $\beta_i \in \{\mathcal{U}, \mathcal{B}\}$ (for $i \in \mathcal{I}_2$), the joint distribution of the states of the corresponding RUs is not affected when the RUs are translated by a count of $k \in \mathbb{Z}$ units:

$$\mathbb{P}\left[\left(\bigcap_{i \in \mathcal{I}_1} T_i^t = \alpha_i\right) \cap \left(\bigcap_{i \in \mathcal{I}_2} C_i^t = \beta_i\right)\right] = \mathbb{P}\left[\left(\bigcap_{i \in \mathcal{I}_1} T_{i+k}^t = \alpha_i\right) \cap \left(\bigcap_{i \in \mathcal{I}_2} C_{i+k}^t = \beta_i\right)\right].$$

Similarly to Sec. A.2, we consider a finite time increment $\Delta t$ and we write:

$$\pi^{\alpha\beta\delta,\eta}(t+\Delta t) \overset{\Delta t \to 0}{\sim} \mathbb{P}\left[T_{i-1}^{t+\Delta t} = \alpha | (T_{i-1}, T_i, T_{i+1})^t = (\overline{\alpha}, \beta, \delta), C_i^t = \eta\right] \pi^{\overline{\alpha}\beta\delta,\eta}(t)$$
$$+ \mathbb{P}\left[T_i^{t+\Delta t} = \beta | (T_{i-1}, T_i, T_{i+1})^t = (\alpha, \overline{\beta}, \delta), C_i^t = \eta\right] \pi^{\alpha\overline{\beta}\delta,\eta}(t)$$
$$+ \mathbb{P}\left[T_{i+1}^{t+\Delta t} = \delta | (T_{i-1}, T_i, T_{i+1})^t = (\alpha, \beta, \overline{\delta}), C_i^t = \eta\right] \pi^{\alpha\beta\overline{\delta},\eta}(t)$$
$$+ \mathbb{P}\left[C_i^{t+\Delta t} = \eta | (T_{i-1}, T_i, T_{i+1})^t = (\alpha, \beta, \delta), C_i^t = \overline{\eta}\right] \pi^{\alpha\beta\delta,\overline{\eta}}(t)$$
$$+ \mathbb{P}\left[(T_{i-1}, T_i, T_{i+1})^{t+\Delta t} = (\alpha, \beta, \delta), C_i^{t+\Delta t} = \eta | \right.$$
$$\left. (T_{i-1}, T_i, T_{i+1})^t = (\alpha, \beta, \delta), C_i^t = \eta\right] \pi^{\alpha\beta\delta,\eta}(t),$$

where, by definition of the transition rates, it holds:

$$\mathbb{P}\left[T_i^{t+\Delta t} = \beta | (T_{i-1}, T_i, T_{i+1})^t = (\alpha, \overline{\beta}, \delta), C_i^t = \eta\right] \overset{\Delta t \to 0}{\sim} k_T^{\overline{\beta}\beta|\alpha \cdot \delta, \eta} \Delta t,$$

and

$$\mathbb{P}\left[C_i^{t+\Delta t} = \eta | (T_{i-1}, T_i, T_{i+1})^t = (\alpha, \beta, \delta), C_i^t = \overline{\eta}\right] \overset{\Delta t \to 0}{\sim} k_C^{\overline{\eta}\eta|\beta} \Delta t.$$

By adopting assumption (H4), Prop. 1 for $A = (T_{i-1}^{t+\Delta t} = \eta)$, $B = (T_{i+1}^t = \delta, C_i^t = \eta)$, $C = ((T_{i-1}, T_i)^t = (\alpha, \beta))$ and $\mathcal{D} = \{(T_{i-2}^t = \xi, C_{i-1}^t = \zeta)\}_{\xi, \zeta}$ leads to:

$$\mathbb{P}\left[T_{i-1}^{t+\Delta t} = \alpha | (T_{i-1}, T_i, T_{i+1})^t = (\overline{\alpha}, \beta, \delta), C_i^t = \eta\right]$$
$$= \frac{\sum_{\xi, \zeta} \mathbb{P}\left[T_{i-1}^{t+\Delta t} = \alpha | (T_{i-2}, T_{i-1}, T_i)^t = (\xi, \overline{\alpha}, \beta), C_{i-1}^t = \zeta\right] \pi_{i-1}^{\xi\overline{\alpha}\beta,\zeta}(t)}{\sum_{\xi, \zeta} \pi_{i-1}^{\xi\overline{\alpha}\beta,\zeta}(t)}$$
$$\overset{\Delta t \to 0}{\sim} \widetilde{k}_T^{\overline{\alpha}\alpha|\circ \cdot \beta, \circ} \Delta t.$$

In conclusion, by letting $\Delta t \to 0$, we get Eq. (15).

## A.4 Derivation of Eqs. (7) and (17)

We start with a remark on Ass. (H3).

*Remark* 1. Assumption (H3) states that, whenever a MH can bind to a given BS, it cannot be involved in a XB with another BS. Suppose that the support of $f$ is contained in the interval $[x_1, x_1 + h]$. Then, this is equivalent to say that, if $d_{ij} \in [x_1, x_1 + h]$, the XBs between the couples $(i-1, j)$ and $(i+1, j)$, which feature displacements $d_{ij} - D_A$ and $d_{ij} + D_A$ respectively, cannot exist. This condition is automatically fulfilled if XBs are present only for displacements in the interval $(-D_A + x_1 + h, D_A + x_1)$, which has width $2D_A - h$. The interval consists in the support of $f$, with width $h$, surrounded by two bands of width $D_A - h$. Consider now the following condition:

$$f_{\mathcal{P}}^i(d_{ij}(t), v(t)) \neq 0 \implies M_k \neq i \quad \forall k \neq j. \tag{H3-bis}$$

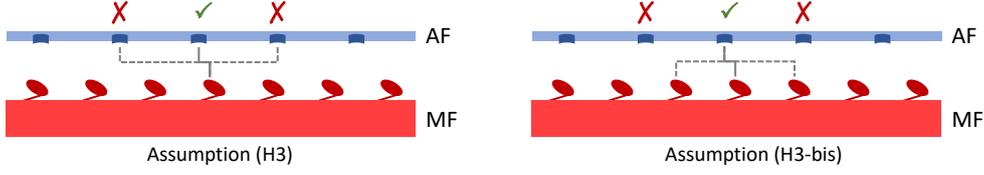

Figure A-1: Representation of assumptions (H3)-(H3-bis). According to assumption (H3) (respectively, assumption (H3-bis)) when a BS-MH pair is within the XB formation range, then the adjacent BSs (respectively, MHs) cannot be bound to the considered MH (respectively, BS).

Assumption (H3-bis) states that, whenever a BS lies within the attachment range of a given MH, it cannot be involved in a XB with another MH. By similar considerations as above, it turns out that this hypothesis is satisfied if XBs are present only in the range $(-D_M + x_1 + h, D_M + x_1)$. Since $D_M > D_A$, assumption (H3) is stronger than (H3-bis). Assumptions (H3)-(H3-bis) allow to decouple the dynamics of the different units. Their validity is justified when the shortening velocity is relatively small, whereas, for large velocities, the XB displacements may be convected outside the region $(-D_A + x_1 + h, D_A + x_1)$. Figure A-1 provides a visual representation of assumptions (H3)-(H3-bis).

We recall that we have defined $d_{ij}(t)$ as the distance between the $i$-th actin BS and the $j$-th MH at time $t$. Since the myofilaments mutually slide with velocity $v_{\text{hs}}(t) = -\frac{d}{dt}SL(t)/2$, we have, for some constant $d_0$:

$$d_{ij}(t) = D_A i - D_M j + \frac{SL(t)}{2} - d_0,$$

$D_A$ and $D_M$ being the distance between two consecutive BSs and MHs, respectively. In order to account for the imperfections in the sarcomere lattice, we consider the value of $d_0$ as a random variable rather than a constant. Hence, we assume that, given a BS in front of the MF, the probability that the closest MH is located at distance $x$ is uniform for $x \in [0, D_M)$. We denote by $\rho_M := \mathbb{f}\left[\exists j \in \mathbb{Z} : d_{ij}^t = x\right] = D_M^{-1}$ the MH linear density, that is:

$$\mathbb{P}\left[\exists j \in \mathbb{Z} : d_{ij}^t = x \in (a, b)\right] = \int_a^b \mathbb{f}\left[\exists j \in \mathbb{Z} : d_{ij}^t = x\right] dx = \rho_M |b - a|.$$

Let us consider now the variable $n_{i,\mathcal{P}}(x, t)$ (similar calculations can be carried out for $n_{i,\mathcal{N}}(x, t)$). We have:

$$\mathbb{f}\left[(Z_i, T_i)^{t+\Delta t} = (x - v_{\text{hs}}(t)\Delta t, \mathcal{P})\right] \overset{\Delta t \to 0}{\sim}$$
$$\mathbb{f}\left[(Z_i, T_i)^{t+\Delta t} = (x - v_{\text{hs}}(t)\Delta t, \mathcal{P})|(Z_i, T_i)^t = (\emptyset, \mathcal{P})\right]\mathbb{P}\left[(Z_i, T_i)^t = (\emptyset, \mathcal{P})\right]$$
$$+\mathbb{f}\left[(Z_i, T_i)^{t+\Delta t} = (x - v_{\text{hs}}(t)\Delta t, \mathcal{P})|(Z_i, T_i)^t = (x, \mathcal{N})\right]\mathbb{f}\left[(Z_i, T_i)^t = (x, \mathcal{N})\right]$$
$$+\mathbb{f}\left[(Z_i, T_i)^{t+\Delta t} = (x - v_{\text{hs}}(t)\Delta t, \mathcal{P})|(Z_i, T_i)^t = (x, \mathcal{P})\right]\mathbb{f}\left[(Z_i, T_i)^t = (x, \mathcal{P})\right].$$

Thanks to Prop. 1, by taking $A = ((Z_i, T_i)^{t+\Delta t} = (x - v_{\text{hs}}(t)\Delta t, \mathcal{P}))$, $B = (Z_i^t = x)$, $C = (T_i^t = \mathcal{N})$ and $\mathcal{D} = \{(T_{i-1}, T_{i+1}, C_i)^t = (\alpha, \eta, \delta)\}_{\alpha, \eta, \delta}$, assumption (H2) leads to

$$\mathbb{P}\left[(Z_i, T_i)^{t+\Delta t} = (x - v_{\text{hs}}(t)\Delta t, \mathcal{P}) | (Z_i, T_i)^t = (x, \mathcal{N})\right] \overset{\Delta t \to 0}{\sim} \widetilde{k}_{T,i}^{\mathcal{NP}} \Delta t.$$

where we have defined:

$$\widetilde{k}_{T,i}^{\mathcal{NP}} := \frac{\sum_{\alpha, \eta, \delta} k_{T,i}^{\mathcal{NP}|\alpha \cdot \eta, \delta} \mathbb{P}\left[(T_{i-1}, T_i, T_{i+1}, C_i)^t = (\alpha, \mathcal{N}, \eta, \delta)\right]}{\mathbb{P}\left[T_i^t = \mathcal{N}\right]},$$

$$\widetilde{k}_{T,i}^{\mathcal{PN}} := \frac{\sum_{\alpha, \eta, \delta} k_{T,i}^{\mathcal{PN}|\alpha \cdot \eta, \delta} \mathbb{P}\left[(T_{i-1}, T_i, T_{i+1}, C_i)^t = (\alpha, \mathcal{P}, \eta, \delta)\right]}{\mathbb{P}\left[T_i^t = \mathcal{P}\right]}.$$

We notice that the transition rates $\widetilde{k}_{T,i}^{\mathcal{NP}}$ and $\widetilde{k}_{T,i}^{\mathcal{PN}}$ can be obtained from the variables $\pi_i^{\alpha\beta\delta,\vartheta\eta\lambda}$ as in Eq. (8). Moreover, we have:

$$\mathbb{P}\left[(Z_i, T_i)^{t+\Delta t} = (x - v_{\text{hs}}(t)\Delta t, \mathcal{P}) | (Z_i, T_i)^t = (x, \mathcal{P})\right]$$
$$\overset{\Delta t \to 0}{\sim} 1 - \mathbb{P}\left[(Z_i, T_i)^{t+\Delta t} = (\emptyset, \mathcal{P}) | (Z_i, T_i)^t = (x, \mathcal{P})\right]$$
$$- \mathbb{P}\left[(Z_i, T_i)^{t+\Delta t} = (x - v_{\text{hs}}(t)\Delta t, \mathcal{N}) | (Z_i, T_i)^t = (x, \mathcal{P})\right]$$
$$\overset{\Delta t \to 0}{\sim} 1 - \Delta t \left(g_{\mathcal{P}}^i(x, v(t)) - \widetilde{k}_{T,i}^{\mathcal{PN}}\right),$$

where we have applied once again assumption (H2). Concerning the XB formation term, we have:

$$(F) := \mathbb{P}\left[(Z_i, T_i)^{t+\Delta t} = (x - v_{\text{hs}}(t)\Delta t, \mathcal{P}) | (Z_i, T_i)^t = (\emptyset, \mathcal{P})\right] \mathbb{P}\left[(Z_i, T_i)^t = (\emptyset, \mathcal{P})\right]$$
$$= \mathbb{P}\left[(Z_i, T_i)^{t+\Delta t} = (x - v_{\text{hs}}(t)\Delta t, \mathcal{P}), (Z_i, T_i)^t = (\emptyset, \mathcal{P})\right]$$
$$= \mathbb{P}\left[(Z_i, T_i)^{t+\Delta t} = (x - v_{\text{hs}}(t)\Delta t, \mathcal{P}), (Z_i, T_i)^t = (\emptyset, \mathcal{P}), \exists j \in \mathcal{I}_M \colon d_{ij}^t = x, M_j^t = 0\right]$$
$$+ \mathbb{P}\left[(Z_i, T_i)^{t+\Delta t} = (x - v_{\text{hs}}(t)\Delta t, \mathcal{P}), (Z_i, T_i)^t = (\emptyset, \mathcal{P}), \exists j \in \mathcal{I}_M \colon d_{ij}^t = x, M_j^t \neq 0\right]$$
$$+ \mathbb{P}\left[(Z_i, T_i)^{t+\Delta t} = (x - v_{\text{hs}}(t)\Delta t, \mathcal{P}), (Z_i, T_i)^t = (\emptyset, \mathcal{P}), \exists j \in \mathbb{Z} \setminus \mathcal{I}_M \colon d_{ij}^t = x\right].$$

The last two terms are at least of second order in $\Delta t$ for $\Delta t \to 0$, while the first term gives:

$$\mathbb{P}\left[(Z_i, T_i)^{t+\Delta t} = (x - v_{\text{hs}}(t)\Delta t, \mathcal{P}), (Z_i, T_i)^t = (\emptyset, \mathcal{P}), \exists j \in \mathcal{I}_M \colon d_{ij}^t = x, M_j^t = 0\right]$$
$$= \mathbb{P}\left[(Z_i, T_i)^{t+\Delta t} = (x - v_{\text{hs}}(t)\Delta t, \mathcal{P}) | (Z_i, T_i)^t = (\emptyset, \mathcal{P}), \exists j \in \mathcal{I}_M \colon d_{ij}^t = x, M_j^t = 0\right]$$
$$\mathbb{P}\left[(Z_i, T_i)^t = (\emptyset, \mathcal{P}), \exists j \in \mathbb{Z} \colon d_{ij}^t = x, M_j^t = 0\right]$$
$$\overset{\Delta t \to 0}{\sim} f_{\mathcal{P}}^i(x, v_{\text{hs}}(t)) \mathbb{P}\left[(Z_i, T_i)^t = (\emptyset, \mathcal{P}), \exists j \in \mathbb{Z} \colon d_{ij}^t = x, M_j^t = 0\right] \Delta t;$$

the remaining two terms are null. Thus:

$$(F) \sim f_{\mathcal{P}}^i(x, v(t)) \Delta t \, \mathbb{P}\left[(Z_i, T_i)^t = (\emptyset, \mathcal{P}), \exists j \in \mathcal{I}_M \colon d_{ij}^t = x, M_j^t = 0\right].$$

By assumption (H3), for any $i$ and $x$ such that $f_{\mathcal{P}}^i(x, v(t)) \neq 0$, the event $(M_j^t = 0)$ for $j$ s.t. $d_{ij}^t = x$ implies the event $(Z_i^t = \emptyset)$, thus:

$$\mathbb{P}\left[(Z_i, T_i)^t = (\emptyset, \mathcal{P}), \exists j \in \mathcal{I}_M \colon d_{ij}^t = x, M_j^t = 0\right]$$
$$= \mathbb{P}\left[(Z_i, T_i)^t = (\emptyset, \mathcal{P}), \exists j \in \mathcal{I}_M \colon d_{ij}^t = x\right]$$
$$= (\mathbb{P}\left[T_i^t = \mathcal{P}, \exists j \in \mathcal{I}_M \colon d_{ij}^t = x\right]$$
$$- \sum_k \mathbb{P}\left[(Z_i, T_i)^t = (x + kD_M, \mathcal{P}), \exists j \in \mathcal{I}_M \colon d_{ij}^t = x\right]),$$

since a BS can be only attached with displacements that are multiple of $D_M$. Moreover, we recall that the RU dynamics is independent of the interaction with XBs and thus of $d_0$ (see Sec. 3.2.1) and that for $i$ and $x$ such that $f^i_{\mathcal{P}}(x, v(t)) \neq 0$ the events ($\exists j \in \mathcal{I}_M: d^t_{ij} = x$) and ($\exists j \in \mathbb{Z}: d^t_{ij} = x$) coincide. Therefore, we have (on the support of $f^i_{\mathcal{P}}$):

$$\mathbb{f}\left[T^t_i = \mathcal{P}, \exists j \in \mathcal{I}_M\, d^t_{ij} = x\right] = \mathbb{P}\left[T^t_i = \mathcal{P}\right] \mathbb{f}\left[\exists j \in \mathbb{Z}: d^t_{ij} = x\right].$$

In addition, since $(Z_i = x + kD_M)$ implies $(\exists j \in \mathbb{Z}: d^t_{ij} = x)$, on the support of $f^i_{\mathcal{P}}$ it holds true:

$$\mathbb{f}\left[(Z_i, T_i)^t = (x + kD_M, \mathcal{P}), \exists j \in \mathcal{I}_M: d^t_{ij} = x\right] = \mathbb{f}\left[(Z_i, T_i)^t = (x + kD_M, \mathcal{P})\right].$$

Since assumption (H3) implies (H3-bis), the unique nonzero term of the sum is $k = 0$ and thus:

$$(F) \sim = f^i_{\mathcal{P}}(x, v(t))\Delta t(\mathbb{P}\left[T^t_i = \mathcal{P}\right]\mathbb{f}\left[\exists j \in \mathbb{Z}: d^t_{ij} = x\right] - \mathbb{f}\left[(Z_i, T_i)^t = (x, \mathcal{P})\right]).$$

Finally, we divide everything by $\Delta t$ we let $\Delta t \to 0$ and we observe that:

$$\frac{n_{i,\mathcal{P}}(x - v_{\text{hs}}(t)\Delta t, t + \Delta t) - n_{i,\mathcal{P}}(x, t)}{\Delta t}$$
$$= \frac{n_{i,\mathcal{P}}(x - v_{\text{hs}}(t)\Delta t, t + \Delta t) - n_{i,\mathcal{P}}(x - v_{\text{hs}}(t)\Delta t, t)}{\Delta t}$$
$$+ \frac{n_{i,\mathcal{P}}(x - v_{\text{hs}}(t)\Delta t, t) - n_{i,\mathcal{P}}(x, t)}{\Delta t\, v_{\text{hs}}(t)} v_{\text{hs}}(t)$$
$$\to \frac{\partial n_{i,\mathcal{P}}}{\partial t}(x, t) - v_{\text{hs}}(t)\frac{\partial n_{i,\mathcal{P}}}{\partial x}(x, t).$$

We get in such a way Eq. (7). Moreover, the expected value of the force exerted by the whole half filament is given by:

$$F_{\text{hf}}(t) = \sum_i \int_{-\infty}^{+\infty} F_{\text{XB}}(x) \mathbb{f}\left[Z^t_i = x\right] dx$$
$$= \sum_i \int_{-\infty}^{+\infty} F_{\text{XB}}(x) \left(\mathbb{f}\left[Z^t_i = x, T^t_i = \mathcal{P}\right] + \mathbb{f}\left[Z^t_i = x, T^t_i = \mathcal{N}\right]\right) dx$$
$$= \sum_i \int_{-\infty}^{+\infty} F_{\text{XB}}(x) \left(n_{i,\mathcal{P}}(x, t) + n_{i,\mathcal{N}}(x, t)\right) dx.$$

On the other hand, Eq. (17) can be derived similarly to Eq. (7), by dropping the dependence on the RU index $i$.

## A.5 Derivation of Eqs. (11) and (21)

By following [111], we multiply Eq. (7) by $(\frac{x}{SL_0/2})^p$, for $p = 0, 1$, and we integrate with respect to $x$ over the real line. Thanks to the fact that, for $x \to \pm\infty$, the distributions $n_{i,\alpha}$ are definitively equal to zero, for $\alpha \in \{\mathcal{N}, \mathcal{P}\}$ and for $i \in \mathcal{I}_A$, the convective terms give raise to the following terms. For $p = 0$, we have:

$$\int_{-\infty}^{+\infty} v_{\text{hs}} \frac{\partial n_{i,\alpha}}{\partial x} dx = [n_{i,\alpha}]_{-\infty}^{+\infty} = 0.$$

On the other hand, for $p = 1$, we have:

$$\int_{-\infty}^{+\infty} \frac{x}{SL_0/2} v_{\text{hs}} \frac{\partial n_{i,\alpha}}{\partial x} dx = -\int_{-\infty}^{+\infty} \frac{v_{\text{hs}}}{SL_0/2} n_{i,\mathcal{P}} dx + v_{\text{hs}} \left[ \frac{x}{SL_0/2} n_{i,\alpha} \right]_{-\infty}^{+\infty} = -v \mu_{i,\alpha}^0(t).$$

Hence, simple calculations lead to Eqs. (11) and (21).

## B Parameters calibration

To calibrate the parameters of the proposed models, we will restrict ourselves to experimental measurements from the literature coming from intact cardiac cells since the skinning procedure alters in a significant (and only partially understood) manner the activation and force generation dynamics [4, 55, 102]. Moreover, thanks to the technique of flura-2 fluorescence, it is nowadays possible to measure the intracellular calcium concentration without depriving the cell of its membrane, and it is also possible to inhibit the sarcoplasmic reticulum calcium uptake by cyclopiazonic acid, so that the calcium level can be controlled without the need of skinning the cells [101, 102].

### B.1 Calibration of the XBs rates

In [82] we have shown that the parameters of the distribution-moments equations describing the XB dynamics can be calibrated starting from five quantities, that are described in the following. Under isometric conditions, we consider the isometric tension $T_a^{\text{iso}} := a_{\text{XB}} \mu^1$ and the fraction of attached XBs $\mu_{\text{iso}}^0 := \mu^0$, where $\mu^0$ and $\mu^1$ are the steady-state solution for $v = 0$. The force-velocity relationship is characterized by the maximum shortening velocity $v^{\max}$, and by $v^0$, its intersection with the axis $T_a = 0$ of the tangent of the curve in isometric conditions, defined as:

$$v^0 := -\left( \left. \frac{\partial \overline{T}_a(v)/T_a^{\text{iso}}}{\partial v} \right|_{v=0} \right)^{-1},$$

where $\overline{T}_a(v)$ denotes the steady-state tension for velocity $v$. Finally, as discussed in [82], the response to fast inputs is characterized, in the small-velocity regime, by the normalized slope of the tension-elongation curve after a fast step. Such quantity is defined as follows: by applying a step in length $\Delta L$ to an isometrically contracted muscle in a small time interval $\Delta t$, we define by $T_a(\Delta t)$ the tension recorded after the step is applied, and we define the normalized stiffness as:

$$\tilde{k}_2 := -\left. \frac{\partial T_a(\Delta t)/T_a^{\text{iso}}}{\partial \Delta L} \right|_{\Delta L=0}.$$

As discussed in [82], when the small-velocity regimes is considered, the quantity $\tilde{k}_2$ corresponds to the slope of the $T_2$-$L_2$ relationship.

In conclusion, as shown in [82], by acting on the five parameters $\mu_{f_\mathcal{P}}^0$, $\mu_{f_\mathcal{P}}^1$, $r_0$, $\alpha$ and $a_{\text{XB}}$ we can fit experimentally measured values concerning the isometric solution ($T_a^{\text{iso}}$ and $\mu_{\text{iso}}^0$), the force-velocity relationship ($v^{\max}$ and $v^0$) and the fast transients response ($\tilde{k}_2$). All the above mentioned experimental setups are such that the thin filament activation machinery can be considered in steady-state. Indeed, $[\text{Ca}^{2+}]_i$ is constant in all the cases and, concerning $SL$: it is also constant under isometric conditions; constant shortening experiments are typically performed in the plateau region of the force-length relationship, and thus the effect of changes in $SL$ is irrelevant; fast transient experiments are carried out at a time-scale such that the activation variables can be considered constant, since they are characterized by a much slower dynamics [7, 52]. Therefore, in these cases, the values of $P_i$, $\tilde{k}_{T,i}^{\mathcal{PN}}$ and $\tilde{k}_{T,i}^{\mathcal{NP}}$ can be considered as fixed in Eq. (11) (and similarly in Eq. (21)).

| Parameter | Value | Units | Reference |
|---|---|---|---|
| $T_\mathrm{a}^\mathrm{iso}$ | 120 | kPa | [101] |
| $\mu_\mathrm{iso}^0$ | 0.22 | - | [9] |
| $v^\mathrm{max}$ | 8 | $\mathrm{s}^{-1}$ | [12] |
| $v^0$ | 2 | $\mathrm{s}^{-1}$ | [12] |
| $\tilde{k}_2$ | 66 | - | [12] |

Table B-1: List of the experimental data used for model calibration.

We notice that, while for the models considered in [82] the relationship between the five parameters and the five experimentally measured values can be analytically inverted, in this case we find the values of the parameters with a numerical strategy. Specifically, to find the steady-state solution we solve Eq. (11) by setting to zero the time derivative terms; we consider the exact solution after the fast transient (the linear ODE system (11) can be solved analytically); we approximate the derivative appearing in the definition of $v^0$ and $\tilde{k}_2$ by finite differences. Finally, we solve, for the five parameters, the nonlinear system of equations linking the five measured values with the parameters themselves. With this aim we employ the Newton-Raphson method, by approximating the Jacobian matrix by means of finite differences.

The experimental data used to calibrate the model are reported in Table B-1, together with a reference to the source in literature. As we mentioned at the beginning of Sec. 4, we employ data coming from room-temperature intact cardiac rat cell, apart from $\mu_\mathrm{iso}^0$ (acquired from skeletal frog muscle), for which we did not find measurements from cardiac muscles. However, as shown in [82], this variable only affects the value of the microscopic variables (i.e. $\mu_{i,\alpha}^p$), but not that of the predicted active tension $T_\mathrm{a}$.

We notice that the constants $v^\mathrm{max}$, $v^0$ and $\tilde{k}_2$ are normalized with respect to $T_\mathrm{a}^\mathrm{iso}$ and are thus valid for a wide range of activation levels (see Sec. 2.2). Conversely, the value of $T_\mathrm{a}^\mathrm{iso}$ is associated with a $SL$ in the plateau region and to saturating calcium concentration. Therefore, when we calibrate the parameters we set $[\mathrm{Ca}^{2+}]_i = 10\,\mu\mathrm{M}$ and $SL = 2.2\,\mu\mathrm{m}$.

## B.2 Calibration of the RUs rates (steady-state)

The steady-state characterization of the muscle tissue activation is represented by the force-calcium and force-length relationships (see Sec. 2.1), whose main features are the behavior for $SL$ in the plateau region (characterized by the tension for saturating calcium $T_\mathrm{a}^\mathrm{iso}$, the calcium sensitivity $\mathrm{EC}_{50}$ and the cooperativity coefficient $n_H$) and the effect of $SL$ (on the saturating tension $T_\mathrm{a}^\mathrm{iso}$ and on the calcium sensitivity $\mathrm{EC}_{50}$).

The tension for saturating calcium concentrations $T_\mathrm{a}^\mathrm{iso}$ in the plateau region of $SL$ is automatically fitted, thanks to Sec. B.1. The effect of $k_\mathrm{d}$ is that of shifting the force-calcium curves with respect to the $\log[\mathrm{Ca}^{2+}]_i$ axes since it only appears in combination with $[\mathrm{Ca}^{2+}]_i$ in the model equations. Therefore, the value of $k_\mathrm{d}$ can be easily calibrated to match the experimental data as it only affects $\mathrm{EC}_{50}$. The effect of $\gamma$, on the other hand, is that of tuning the amount of cooperativity (indeed, it acts on $n_H$). The role of the remaining parameters ($Q$ and $\mu$) is more involved and cannot be easily decoupled, as they affect the cooperativity, calcium sensitivity, and the asymmetry of the force-calcium relationship below and above EC$_{50}$ [90]. Moreover, in the SE-ODE case, they also act on the $SL$-driven regulation on calcium sensitivity.

In the following we set $\mu = 10$, coherently with the fact that the experimentally measured dissociation rate of Tn from calcium varies of one order of magnitude in different combinations [67]. For the SE-ODE model, we set $\gamma$, $Q$ and $k_\mathrm{d}$, to fit the steady-state force-calcium measurements of [101] (referred to the two different values of $SL$ of 1.85 and 2.15 µm) from intact rat cardiac cells at room temperature.

## B.3 Calibration of the RUs rates (kinetics)

We consider the isometric twitches of intact rat cardiac muscle fibers reported in [48] for different values of $SL$, and with $[\mathrm{Ca}^{2+}]_\mathrm{o} = 1.0\,\mathrm{mM}$. Since the corresponding calcium transients are not reported, we consider the calcium transient reported in [49] for the same muscle preparation. As the reported trace is much affected by noise, we fit it with the following idealized transient [107]:

$$[\mathrm{Ca}^{2+}]_\mathrm{i}(t) = \begin{cases} c_0 & t < t_0^c, \\ c_0 + \frac{c_\mathrm{max} - c_0}{\beta}\left[e^{-\frac{t-t_0^c}{\tau_1^c}} - e^{-\frac{t-t_0^c}{\tau_2^c}}\right] & t \geq t_0^c, \end{cases}$$

where

$$\beta = \left(\frac{\tau_1^c}{\tau_2^c}\right)^{-\left(\frac{\tau_1^c}{\tau_2^c}-1\right)^{-1}} - \left(\frac{\tau_1^c}{\tau_2^c}\right)^{-\left(1-\frac{\tau_2^c}{\tau_1^c}\right)^{-1}},$$

with the constants values $c_\mathrm{max} = 1.35\,\mathrm{\mu M}$, $t_0^c = 0.05\,\mathrm{s}$, $\tau_1^c = 0.02\,\mathrm{s}$, $\tau_2^c = 0.19\,\mathrm{s}$.

Then, we sample the candidate parameters space $(k_\mathrm{basic}, k_\mathrm{off}) \in [0, 80\,\mathrm{s}^{-1}] \times [0, 300\,\mathrm{s}^{-1}]$ by a MC strategy, for each parameters value we compute the tension transients corresponding to $SL = 1.90$, $2.05$ and $2.20\,\mathrm{\mu m}$ and we compare them with the experimental measurements from [48]. We consider the following discrepancy metrics, where $T_\mathrm{a}^{i,\mathrm{mod}}(t)$ denotes the tension predicted by the model for the $i$-th value of $SL$ and $T_\mathrm{a}^{i,\mathrm{exp}}(t)$ denotes the experimentally measured one:

$$E_{L^2} := \sqrt{\sum_{i=1}^{3} \int_0^T \left|T_\mathrm{a}^{i,\mathrm{mod}}(t) - T_\mathrm{a}^{i,\mathrm{exp}}(t)\right|^2 dt},$$

$$E_\mathrm{peak} := \sqrt{\sum_{i=1}^{3} \left|\sup_{t \in [0,T]} T_\mathrm{a}^{i,\mathrm{mod}}(t) - \sup_{t \in [0,T]} T_\mathrm{a}^{i,\mathrm{exp}}(t)\right|^2}.$$

The first metric coincides with the $L^2$ error over time, while the second one evaluates the error of the predicted force peak. We notice indeed that the parameters $k_\mathrm{basic}$ and $k_\mathrm{off}$ also determine the force peak attained during a sarcomere twitch: the most rapid the tissue activation is, the more the force-calcium curve stays close to the steady-state curve and thus it reaches higher force values before the relaxation begins. As criterion to select the best parameters values, we consider as overall metric a weighted combination between the two, given by $E_\mathrm{tot} = (E_{L^2}^2 + w_\mathrm{peak}^2 E_\mathrm{peak}^2)^{1/2}$, where we set $w_\mathrm{peak} = 5$.

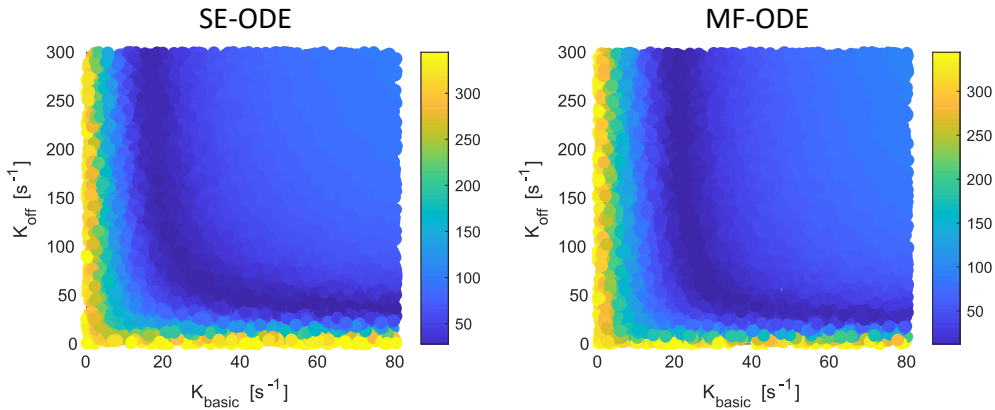

Figure B-1: Discrepancy metric $E_{\text{tot}}$ in the parameters space for intact, room-temperature rat cells, obtained with the SE-ODE model (left) and with the MF-ODE model (right).

The obtained values of the discrepancy metric $E_{\text{tot}}$ in the parameters space for both the SE-ODE and the MF-ODE models are reported in Fig. B-1.

We notice that the level curves do not clearly identify an optimal pair $(k_{\text{basic}}, k_{\text{off}})$, rather these exhibit a wide region in the parameters space producing very similar results. Given the uncertainty in the measurements of both force and, mostly, calcium, it makes no sense to select the best parameters by blindly selecting the pair that realizes the smaller discrepancy from experimental results. Therefore, we supplement the results of Fig. B-1 with direct measurements of calcium binding rates to Tn, showing that $k_{\text{on}} = k_{\text{off}}/k_{\text{d}}$ lies between 50 and 200 $\mu\text{M}^{-1}\,\text{s}^{-1}$ [67]. On this basis, we restrict the region of candidate values and we select the parameters reported in Tab. 3.

The predicted isometric twitches obtained with the selected values of the parameters are compared with the experimental data in Fig. 15. We notice here that the MF-ODE model accomplishes a better fit of experimental data than the SE-ODE model. This is an effect of the phenomenological tuning of $k_{\text{d}}$ of Eq. (29), that allows for a significant increase of calcium sensitivity and, consequently, of twitch duration, for higher values of $SL$. Nonetheless, also the SE-ODE model predicts, even if to a lower extent, the experimentally observed prolongation of twitches at higher $SL$, without any phenomenological tuning of the calcium sensitivity.

We notice that there is room for improvement in the calibration of the kinetic parameters $k_{\text{basic}}$ and $k_{\text{off}}$, which could be better constrained in presence of more abundant and more reliable experimental data and when a deeper understanding on the determinants of the kinetics of activation and relaxation will be available. Nevertheless, the calibration of $k_{\text{basic}}$ and $k_{\text{off}}$ for the rat model does not affect the quality of the human model, since those two parameters are the only ones to be completely re-calibrated for the human model.

### B.4 Human model at body temperature

In order to adapt the parameters calibrated from rat data at room temperature to a body-temperature human model, we first focus on the steady state, to reflect a higher calcium sensitivity. For this purpose, we employ the data reported in [61], which, however, refer to skinned cells. In order to estimate the effect of skinning on $k_\mathrm{d}$, we compare the calcium sensitivity measured for room-temperature rat cardiac cells in skinned [27] and intact preparations [101] at $SL = 2.15\,\mathrm{\mu m}$ and we assume that the same relationship holds for skinned versus intact, body-temperature human cells. Finally, we rescale the values of $k_\mathrm{d}$ to reflect the estimated change in calcium sensitivity between intact, body-temperature human cells and intact, room-temperature rat cells, obtaining the values reported in Tab. 3. The experimental data used in such procedure are listed in Tab. B-2.

| Species | Temperature | Preparation | $SL$(μm) | $EC_{50}$(μM) | Reference |
|---|---|---|---|---|---|
| Rat | Room | Skinned | 2.15 | 3.51 | [27] |
| Rat | Room | Intact | 2.15 | 0.68 | [101] |
| Human | Body | Skinned | 2.00 | 1.72 | [61] |
| Human | Body | Skinned | 2.20 | 1.56 | [61] |

Table B-2: List of the experimental values used to calibrate the calcium sensitivity for the human models at body temperature.

Since the RUs kinetics may depend on both the species and the temperature, we re-calibrate the parameters $k_\mathrm{off}$ and $k_\mathrm{basic}$ on the base of the kinetic metrics reported in [18] (the data are referred to body-temperature human cells). These metrics include the peak tension $T_\mathrm{a}^\mathrm{peak}$, the time-to-peak $TTP$ (defined as the time separating the beginning of force raise and the tension peak) and the relaxation times $RT_{50}$ and $RT_{95}$ (defined as the time between the tension peak and 50% and 95% of relaxation, respectively). Since, as to the best of our knowledge, detailed calcium transients measurements for intact human cells at body temperature are not currently available, we employ the synthetic calcium transient predicted by the ToR-ORd model [104]. The metrics reported in [18] are referred to different values of $SL$, expressed as fraction of the *optimal sarcomere length* (i.e. the length for which an increase of developed force is compensated by an increase in resting tension), corresponding, according to the authors, to nearly $SL = 2.2\,\mathrm{\mu m}$. For the calibration, we employ the value associated with 95% of the optimal length (i.e. $2.09\,\mathrm{\mu m}$).

Finally, for the calibration of the parameters ruling the XBs cycling, we use the same values of Tab. B-1. Therefore, since the calibration depends on the parameters previously set for the RUs activation, the resulting values of the parameters are slightly different. We provide in Tab. 3 the full list of parameters for both species (room-temperature rat and body-temperature human) and for both models (SE-ODE and MF-ODE).

## C Numerical schemes

We provide details about the numerical schemes employed to approximate the solution of the models proposed in this paper.

### C.1 RU dynamics

Since the equations describing the evolution of the RUs are independent of the variables describing the XB states, their solution can be approximated independently of that of the models describing the XB dynamics. Specifically, we adopt a Forward

Euler scheme with a time step size of $2.5 \cdot 10^{-5}$ s. Further details on the properties of the numerical solution obtained by applying this scheme can be found in [85].

## C.2 XB dynamics

Concerning the equations describing the XB dynamics, we notice that Eqs. (11) and (21) can be written in the following form:

$$\begin{cases} \dot{\mathbf{x}}(t) = \mathbf{A}(t)\mathbf{x}(t) + \mathbf{r}(t) & t \in (0,T], \\ \mathbf{x}(0) = \mathbf{x}_0, \end{cases} \quad (30)$$

where $\mathbf{x}(t)$ is the vector of the variables describing the states of XBs, while $\mathbf{A}(t)$ and $\mathbf{r}(t)$ are respectively a time-dependent matrix and vector, determined by the input $v(t)$ and by the RUs states $\pi_i^{\alpha\beta\delta,\vartheta\eta\lambda}(t)$ (or, for mean-field models, $\pi^{\alpha\beta\delta,\eta}(t)$).

In order to approximate the solution of Eq. (30), we consider a subdivision $0 = t_0 < t_1 < \cdots < t_M = T$ of the time interval $[0,T]$ with time step size $\Delta t$ and we denote by $\mathbf{x}^{(k)} \approx \mathbf{x}(t_k)$ the approximated solution at time $t_k$. Due to the linearity of Eq. (30), we consider the following exponential scheme [42]:

$$\begin{cases} \mathbf{x}^{(0)} = \mathbf{x}_0, \\ \mathbf{x}_\infty^{(k)} = -\mathbf{A}^{-1}(t_k)\mathbf{r}(t_k) & \text{for } k \geq 1, \\ \mathbf{x}^{(k)} = \mathbf{x}_\infty^{(k)} + e^{\Delta t \mathbf{A}(t_k)}(\mathbf{x}^{(k-1)} - \mathbf{x}_\infty^{(k)}) & \text{for } k \geq 1. \end{cases} \quad (31)$$

Due to the implicit nature of the scheme of Eq. (31), that entails better stability properties than the explicit scheme used for the RUs equations [76], we solve it with a larger time step size than the one used for the RUs model ($\Delta t = 1 \cdot 10^{-3}$ s), by a first-order time splitting scheme.